\documentclass[aps,prx,twocolumn,superscriptaddress,floatfix]{revtex4-2}

\usepackage{graphicx}
\usepackage{dcolumn}
\usepackage{bm}
\usepackage{amsmath, amssymb}
\usepackage{xcolor}
\usepackage{comment}
\usepackage{float}
\usepackage{hyperref}
\usepackage{algorithm}
\usepackage{enumitem}
\usepackage{booktabs}
\usepackage{algorithmic}
\usepackage{ragged2e}
\usepackage[most]{tcolorbox}
\hypersetup{
    colorlinks=true,
    linkcolor=blue,
    citecolor=blue,
    urlcolor=blue
}

\begin{document}

\title {Adaptive Quantum Matter: Variational Organization through Ising Agents}

\author{Lakshya Nagpal}
\affiliation{Institute of Mathematical Sciences, CIT Campus, Tharamani 600113}
\affiliation{Pecslab Education and Research}
\email{lakshyan@imsc.res.in}

\author{S. R. Hassan}
\affiliation{Institute of Mathematical Sciences, CIT Campus, Tharamani 600113}
\affiliation{Homi Bhabha National Institute, Training School Complex, Anushakti Nagar, Mumbai 400094, India}
\email{shassan@imsc.res.in}

\date{\today}

\begin{abstract}
The study introduces the Adaptive Quantum Ising Agents (AQIA) framework, a Hamiltonian-based methodology that extends programmable quantum matter into an adaptive domain. Each agent operates as a finite transverse-field Ising subsystem, maintaining internal quantum coherence while interacting through state-dependent feedback channels characterized by reduced observables, such as spin polarization, bond correlation, and internal energy. These informational couplings enable the transformation of a static lattice into a feedback-reconfigurable medium. The effective Hamiltonian generated, which remains Hermitian at each iteration, is resolved self-consistently using a mean-field approximation. Here, the feedback fields are iteratively adjusted to optimize total energy minimization. Numerical investigations identify three distinct regimes: domain formation proximal to the feedback–fluctuation critical point, glass-like frustration due to competing feedback channels, and modular polarization sustained by structured interactions. These phenomena occur independently of geometric embedding, illustrating that informational similarity alone can induce coherent organization. The AQIA framework is adaptable to implementation on superconducting, trapped-ion, or Rydberg platforms, offering a minimalistic model for exploring self-organization and learning in adaptive programmable quantum matter.

\end{abstract}

\maketitle
\section{Introduction}
\label{sec:intro}

Collective organization is a unifying principle across physics, biology, and engineered systems.  
In condensed matter, macroscopic order arises from microscopic couplings—manifesting as magnetism, superconductivity, and correlated screening phenomena~\cite{carr_understanding_2010,sachdev_quantum_2011,cross_pattern_1993}.  
In driven or active media, coordination emerges through local sensing and feedback, producing synchronization and pattern formation without central control~\cite{cross_pattern_1993,pikovsky_synchronization_2001,anderson_more_1972}.  
Across these diverse settings, large-scale coherence arises when many interacting units continually adjust to one another.

In conventional many-body theory, the structure of interactions is externally prescribed:
the Hamiltonian defines fixed couplings set by geometry or control parameters.
This paradigm underlies most equilibrium and non-equilibrium models in quantum materials. ~\cite{nandkishore_many-body_2015,eisert_quantum_2015,browaeys_many-body_2020,eisert_quantum_2015}
For example, in a spin system the Ising Hamiltonian
\begin{equation}
H_{\mathrm{Ising}} = -\sum_{i<j} J_{ij}\, Z_i Z_j - \sum_i h_i Z_i
\end{equation}
contains coupling coefficients $J_{ij}$ that are fixed by lattice geometry or by engineered physical links.
Similarly, in Rydberg-atom simulators the effective potential
\begin{equation}
V_{ij} = \frac{C_6}{|\mathbf{r}_i - \mathbf{r}_j|^6}
\end{equation}
sets the interaction between atoms located at positions $\mathbf{r}_i$ and $\mathbf{r}_j$.
These coefficients $J_{ij}$ and $V_{ij}$ encode real-space interactions that are static during any given experiment.

With the advent of programmable quantum hardware—Rydberg-atom arrays, trapped-ion crystals, and superconducting-qubit lattices—experimenters can now \emph{reprogram} these couplings in situ by tuning optical or microwave control fields~\cite{kjaergaard_superconducting_2020,schutski_adaptive_2020}.
Such systems realize what is broadly known as \emph{programmable quantum matter}, in which the interaction matrix $\{J_{ij}\}$ or $\{V_{ij}\}$ can be externally adjusted to realize different geometries or coupling patterns.
However, the tuning logic remains classical and unidirectional: the device does not modify its own couplings based on the quantum state it produces.

This distinction motivates the search for a Hamiltonian formulation that incorporates feedback as an internal mechanism rather than an external instruction.  
If local observables of a quantum subsystem could influence its future interactions with others, the ensemble would constitute an \emph{adaptive quantum medium}—a self-reconfiguring network whose couplings evolve in response to its own state. ~\cite{altamirano_unitarity_2017,friedman_measurement-induced_2023,schutski_adaptive_2020}
Such an idea parallels artificial-life models in classical complex systems, where agents update their rules through local feedback and collectively develop organized behaviour ~\cite{ventrella_designing_1998}.  
In the quantum context, this requires embedding feedback within the Hamiltonian itself while preserving hermiticity and physical consistency.

\emph{Adaptive Quantum Ising Agents} (AQIA) provide a minimal realization of this concept.  
Each agent is a finite transverse-field Ising subsystem that remains fully quantum internally but interacts with others through couplings determined by reduced observables of their respective ground states—spin polarization, bond correlation, and internal energy.  
These quantities serve as informational summaries that define an emergent interaction network in ``information space.''~\cite{cerezo_variational_2021,parrondo_thermodynamics_2024}
The collective dynamics are governed by an effective Hamiltonian of the form
\begin{equation}
H_{\mathrm{eff}}
   = \sum_i H_i^0
     - \sum_{i<j} K_{ij}(\mathbf{m}_i,\mathbf{m}_j)\, O_i O_j ,
\label{eq:Hschematic}
\end{equation}
where $O_i$ denotes a representative operator of agent $i$, and $K_{ij}$ is a feedback-dependent coupling that evolves with the observable summaries $\mathbf{m}_i$.  
Unlike fixed-geometry lattices, the coupling matrix $\{K_{ij}\}$ is a \emph{variational object} determined self-consistently by the ensemble itself.  
Solving for its stationary configuration corresponds to minimizing the collective ground-state energy of $H_{\mathrm{eff}}$.  
In principle, this optimization could be performed using any advanced many-body method—such as dynamical mean-field theory, tensor-network approaches, or quantum Monte Carlo. ~\cite{ortega_iterative_2000,georges_dynamical_1996,nandkishore_many-body_2015} 
In this work, however, we employ a mean-field approximation to provide a transparent proof of concept, demonstrating that even at this level the feedback-driven system exhibits rich adaptive organization and well-defined emergent phases.

This framework unites two paradigms within a single Hamiltonian language.  
From condensed-matter physics it inherits microscopic consistency and mean-field solvability;  
from adaptive networks and artificial-life theory it adopts distributed feedback and self-organization. ~\cite{kiefer_quantum_2010,altamirano_unitarity_2017,ivanov_tuning_2021}
The resulting model is a geometry-free quantum network where information exchange, rather than spatial distance, defines connectivity.  
Through this coupling of quantum coherence and adaptive feedback, AQIA establishes a foundation for \emph{adaptive programmable matter}:  
quantum systems capable of reorganizing their own interaction landscape to achieve emergent order.  ~\cite{stumpf_more_2022,cross_pattern_1993-1,khaneja_optimal_2005,ventrella_designing_1998}

The remainder of this paper develops this framework in detail.  
Section~\ref{sec:model} formulates the agent-level Hamiltonians and feedback kernels.  
Section~\ref{sec:meanfield} derives the mean-field self-consistency equations used to solve $H_{\mathrm{eff}}$.  
Section~\ref{sec:regimes} presents numerical results exhibiting adaptive criticality, frustration, and modular polarization.  
Section~\ref{sec:finite_size_scaling} connects these behaviours to experimental realizations in feedback-controlled quantum platforms.  
Finally, Section~\ref{sec:conclusion} outlines the broader implications of AQIA as a model of self-organizing quantum matter.

%==========================================================
\section{Model Formulation}
\label{sec:model}
%==========================================================

\subsection{Local agents as quantum patches}
\label{sec:local_agents}

The conceptual framework of Adaptive Quantum Ising Agents (AQIA), introduced in Sec.~\ref{sec:intro}, is now formulated in explicit Hamiltonian form. 
Each agent represents a mesoscopic quantum subsystem—a finite transverse-field Ising model (TFIM) patch—whose internal dynamics remain fully quantum while its couplings to other agents evolve adaptively based on measurable observables of its own ground state. 
Here, the term \emph{adaptive} refers to the feedback dependence of inter-agent couplings on instantaneous local summaries of quantum observables. 
An \emph{ensemble} of such agents constitutes a population of interacting subsystems whose effective connectivity evolves in response to these summaries, forming an \emph{adaptive interaction graph} in information space rather than a fixed spatial lattice.

Formally, agent \(i\) is described by a local TFIM Hamiltonian
\begin{align}
H_i^0(\mathbf{h}_i;\mathbf{J}^{(i)},\Gamma)
  &=
  - \sum_{k=1}^{n} h_{i,k}\, Z_{i,k}
  - \!\!\sum_{\langle k,\ell\rangle\in\mathcal{B}_i}
     J^{(i)}_{k\ell}\, Z_{i,k} Z_{i,\ell}
  - \Gamma \sum_{k=1}^{n} X_{i,k},
\label{eq:Hi0}
\end{align}
where $Z_{i,k}$ and $X_{i,k}$ are Pauli matrices acting on qubit $k$ of agent $i$, 
$\mathbf{h}_i$ and $\mathbf{J}^{(i)}$ denote local fields and intra-agent couplings on the internal bond set $\mathcal{B}_i$, and $\Gamma$ is the transverse field.
We set $\hbar=1$ and measure all energies in units of $\Gamma$.

From the ground (or low-energy) state of $H_i^0$, we extract reduced observables—the local spin polarization, bond correlation, and energy density—
\begin{align}
S_i &= \frac{1}{n}\sum_{k}\langle Z_{i,k}\rangle,\quad
B_i = \frac{1}{|\mathcal{B}_i|}\sum_{(k,\ell)\in\mathcal{B}_i}\langle Z_{i,k}Z_{i,\ell}\rangle,\quad
U_i = \frac{1}{n}\langle H_i^0\rangle,
\label{eq:local_observables}
\end{align}
which together form a compact ``summary vector'' $\mathbf{m}_i=(S_i,B_i,U_i)$ characterizing the internal state of each agent. 
These quantities act as informational order parameters mediating feedback-dependent couplings to the rest of the ensemble.

%----------------------------------------------------------
\subsection{Operator representatives}
\label{sec:operators}
%----------------------------------------------------------

Since $(S_i,B_i,U_i)$ are expectation values, the corresponding operators must be retained to ensure Hermiticity of all feedback updates. 
We define
\begin{align}
\hat S_i &= \frac{1}{n}\sum_{k=1}^{n} Z_{i,k},\qquad
\hat B_i = \frac{1}{|\mathcal{B}_i|}\sum_{(k,\ell)\in\mathcal{B}_i} Z_{i,k}Z_{i,\ell},\qquad
\hat U_i = \frac{1}{n} H_i^0,
\label{eq:operator_reps}
\end{align}
each acting solely on the Hilbert space of agent $i$.  
Inter-agent products such as $\hat S_i\hat S_j$ or $\hat B_i\hat U_j$ generate two-body couplings between agents, while the global mean-field state remains separable.

%----------------------------------------------------------
\subsection{Rule-dependent couplings on an adaptive graph}
\label{sec:couplings}
%----------------------------------------------------------

Interactions between agents are defined on an emergent \emph{similarity graph} whose edges depend on the proximity of their observable summaries. 
For any observable $O\!\in\!\{S,B,U\}$ with ensemble mean $\mu_O$ and standard deviation $\sigma_O$, we introduce the normalized difference
\begin{equation}
\Delta_{ij}^{(O)}=\frac{O_i-O_j}{\sigma_O+\epsilon}, \qquad \epsilon\!\sim\!10^{-6}.
\label{eq:delta_norm}
\end{equation}
To preserve smoothness and differentiability of feedback, the couplings are defined through Gaussian kernels,
\begin{align}
w_{ij}^{(S)} &= S_i S_j\, e^{-\tfrac{1}{2}[\Delta_{ij}^{(S)}]^2},\quad
w_{ij}^{(B)} = B_i B_j\, e^{-\tfrac{1}{2}[\Delta_{ij}^{(B)}]^2},\nonumber\\
w_{ij}^{(U)} &= 
  \frac{\tilde U_i \tilde U_j}{\sigma_U^2}\,
  e^{-\tfrac{1}{2}[\Delta_{ij}^{(U)}]^2},\qquad
\tilde U_i = U_i - \mu_U,
\label{eq:diag_weights}
\end{align}
with cross-channels coupling distinct observables multiplicatively, e.g.
\begin{equation}
w_{ij}^{(SB)}
 = \tfrac{1}{2}(S_i B_j + S_j B_i)
   \exp\!\left[-\tfrac{1}{4}
   \big([\Delta_{ij}^{(S)}]^2 + [\Delta_{ij}^{(B)}]^2\big)\right],
\label{eq:sb_weight}
\end{equation}
and similarly for $SU$ and $BU$ channels.  
All weights are symmetric, $w_{ij}^{(\alpha)}=w_{ji}^{(\alpha)}$, and vanish for $i=j$. 
Such Gaussian-weighted similarity rules are analogous to kernel couplings in adaptive network and Hopfield-type models~\cite{hopfield_1982_1988,khaneja_optimal_2005,newman_finding_2004}, ensuring continuous coupling adaptation while avoiding abrupt topological changes.

%----------------------------------------------------------
\subsection{Effective Hamiltonian and variational energy}
\label{sec:Heff_section}
%----------------------------------------------------------

Given a fixed configuration of summaries $\{\mathbf{m}_i\}$, the adaptive couplings $\{w_{ij}^{(\alpha)}\}$ define an instantaneous many-body Hamiltonian for the entire ensemble:
\begin{equation}
H_{\mathrm{eff}}
   = \sum_i H_i^0
     - \sum_{i<j,\alpha}
       w_{ij}^{(\alpha)}\,
       \hat O_i^{(\alpha)} \hat O_j^{(\alpha)},
\label{eq:Heff}
\end{equation}
where $\hat O_i^{(\alpha)}$ are the operator representatives of observables $\alpha\!\in\!\{S,B,U\}$ acting on agent $i$. 
Equation~\eqref{eq:Heff} serves as a variational generator for the adaptive network: each feedback iteration defines a new Hermitian $H_{\mathrm{eff}}$, whose couplings depend parametrically on the agents’ instantaneous summaries.

For a general many-body state $\rho$ of the ensemble, the corresponding energy expectation is
\begin{equation}
E[\rho;\{\mathbf{m}_i\}]
   = \mathrm{Tr}\!\big(\rho\, H_{\mathrm{eff}}[\{\mathbf{m}_i\}]\big),
\label{eq:energy_expect}
\end{equation}
which may be evaluated using any many-body solver (e.g., exact diagonalization, DMRG, DMFT, or variational ansätze). 
The self-consistent feedback rule simply requires that the observables $\langle \hat O_i^{(\alpha)} \rangle$ used to update the couplings are consistent with the minimizing state of $H_{\mathrm{eff}}$.

%----------------------------------------------------------
\subsection{Mean-field closure and adaptive energy functional}
\label{sec:energy_mf}
%----------------------------------------------------------

To demonstrate the minimal self-consistent mechanism, we adopt a mean-field closure in which the ensemble density matrix factorizes as 
$\rho=\bigotimes_i\rho_i$. 
In this approximation,
\[
\langle \hat O_i^{(\alpha)} \hat O_j^{(\beta)} \rangle
 \approx 
 \langle \hat O_i^{(\alpha)} \rangle
 \langle \hat O_j^{(\beta)} \rangle,
\]
reducing Eq.~\eqref{eq:energy_expect} to an energy functional of the summaries $\mathbf{m}_i=(S_i,B_i,U_i)$:
\begin{equation}
E_{\mathrm{MF}}[\{\mathbf{m}_i\}]
 = \sum_i U_i
   - \!\!\sum_{i<j,\alpha}
     \sum_{\beta\ge\alpha}
     w_{ij}^{(\alpha\beta)}\,
     O_i^{(\alpha)} O_j^{(\beta)}.
\label{eq:energy_functional_mf}
\end{equation}
Here $O_i^{(\alpha)}=\langle\hat O_i^{(\alpha)}\rangle$ represents $\{S_i,B_i,U_i\}$, and $w_{ij}^{(\alpha\beta)}$ compactly encodes both diagonal ($SS$, $BB$, $UU$) and mixed ($SB$, $SU$, $BU$) coupling channels. 
Minimisation of $E_{\mathrm{MF}}$ defines the adaptive equilibrium configuration, with all feedback terms determined self-consistently.

Under mean-field factorisation, inter-agent entanglement is absent; collective phenomena emerge from classical feedback between quantum subsystems rather than genuine quantum correlations

%----------------------------------------------------------
\subsection{Feedback iteration, adaptive regimes, and numerical setup}
\label{sec:meanfield}
%----------------------------------------------------------

At each iteration, the coupling weights $\{w_{ij}^{(\alpha)}\}$ are recomputed from the current summaries $\{\mathbf{m}_i\}$, generating renormalized fields acting on each agent:
\begin{align}
\Phi_i^{S} &= \sum_j \big( w_{ij}^{(S)} S_j + w_{ij}^{(SB)} B_j + w_{ij}^{(SU)} U_j \big), \nonumber\\
\Phi_i^{B} &= \sum_j \big( w_{ij}^{(B)} B_j + w_{ij}^{(SB)} S_j + w_{ij}^{(BU)} U_j \big), \nonumber\\
\Phi_i^{U} &= \sum_j \big( w_{ij}^{(U)} U_j + w_{ij}^{(SU)} S_j + w_{ij}^{(BU)} B_j \big).
\label{eq:renorm_fields}
\end{align}
Each agent then evolves under its instantaneous mean-field Hamiltonian
\begin{equation}
H_i^{\mathrm{MF}}
 = H_i^0
 - \Phi_i^{S}\hat S_i
 - \Phi_i^{B}\hat B_i
 - \Phi_i^{U}\hat U_i,
\label{eq:Hmf}
\end{equation}
which remains Hermitian for fixed $\Phi_i^\alpha$. 
Solving $H_i^{\mathrm{MF}}$ provides updated expectation values $(S_i,B_i,U_i)$, defining the iterative map
\begin{equation}
\mathbf{m}\;\mapsto\;\mathcal{F}[\mathbf{m}],
\qquad
\mathbf{m}=(\mathbf{m}_1,\mathbf{m}_2,\ldots).
\label{eq:self_consistency_map}
\end{equation}
Successive applications of $\mathcal{F}$ drive the ensemble toward a stationary configuration $\mathbf{m}^\star$ that (numerically) minimizes $E_{\mathrm{MF}}[\mathbf{m}]$. 
Although the feedback loop is non-unitary—due to the inclusion of measurement and classical update—the evolution is composed of well-defined Hermitian snapshots, ensuring internal quantum consistency while the total energy is observed to decrease monotonically toward equilibrium.

To explore emergent organization, we solve the adaptive equations for ensembles of disordered initial conditions and average over multiple random realizations of local fields $\{h_i\}$ and couplings $\{J_{k\ell}^{(i)}\}$. 
Three representative parameter regimes capture the principal forms of adaptive order:  
(i) the \emph{critical-balance regime}, with $\langle J\rangle=\langle h\rangle=1$, narrow dispersions $\sigma_J=0.01$, $\sigma_h=0.1$, and $\Gamma=1$, representing competition between order and quantum fluctuations;  
(ii) the \emph{glassy regime}, with broadened random $J$ and $h$, producing frustration and heterogeneous minima; and  
(iii) the \emph{community-polarization regime}, with weaker average coupling $\langle J\rangle=0.5$ and sparse connectivity, yielding modular yet internally coherent clusters.  
These regimes span the adaptive phase space discussed in Sec.~\ref{sec:regimes}.

At convergence, ensemble observables are computed by averaging over $R$ independent \emph{disorder realizations}, each labeled by an index $r=1,\dots,R$, corresponding to distinct draws of local fields $\{h_i^{(r)}\}$ and intra-agent couplings $\{J_{k\ell}^{(i,r)}\}$.  
The Edwards–Anderson order parameter for realization $r$ is defined as
\begin{equation}
q_{\mathrm{EA}}^{(r)} = \frac{1}{N}\sum_{i=1}^{N} \bigl\langle S_i \bigr\rangle_r^2,
\end{equation}
and its ensemble average,
\begin{equation}
\overline{q_{\mathrm{EA}}} = \frac{1}{R}\sum_{r=1}^{R} q_{\mathrm{EA}}^{(r)},
\end{equation}
quantifies the degree of frozen-in local order across the ensemble, distinguishing coherent, glassy, and modular equilibria.  

Fluctuations across realizations are captured by the coefficient of variation,
\begin{equation}
\mathrm{CV}(q_{\mathrm{EA}}) 
  = \frac{
     \sqrt{\frac{1}{R}\sum_{r=1}^{R}\big(q_{\mathrm{EA}}^{(r)}-\overline{q_{\mathrm{EA}}}\big)^2}
     }{\overline{q_{\mathrm{EA}}}},
\end{equation}
where a small $\mathrm{CV}$ indicates convergence toward a single dominant equilibrium, while a large value signals many metastable minima.  

Finally, the modularity $Q$, defined using standard network-theoretic measures,
\begin{equation}
Q = \frac{1}{2W}\sum_{i\neq j}\!\left(w_{ij} - \frac{k_i k_j}{2W}\right)\!\delta_{c_i,c_j},
\end{equation}
characterizes emergent community structure within the adaptive network, with node degrees $k_i=\sum_j w_{ij}$ and total weight $W=\tfrac{1}{2}\sum_{i,j}w_{ij}$.

\newtcolorbox{myalgo}[1]{colback=gray!5!white,colframe=black!50,title=#1}

\begin{myalgo}{AQIA Ensemble Evolution}
\textbf{Initialize:} $\{\mathbf{m}_i^{(0)}\}$ with random disorder fields and couplings.\\
\textbf{Repeat:}
\begin{enumerate}
\item Compute couplings $w_{ij}^{(\alpha)}[\mathbf{m}^{(n)}]$ for all $\alpha \in \{S,B,U\}$.
\item Form renormalized fields $\Phi_i^{(\alpha)}$ via Eq.~\eqref{eq:renorm_fields}.
\item Construct mean-field Hamiltonian $H_i^{\mathrm{MF}}[\Phi_i^{(\alpha)}]$ for each agent $i$.
\item Solve for ground state and update: $\mathbf{m}_i^{(n+1)} \leftarrow \langle \hat{\mathbf{m}}_i \rangle$.
\end{enumerate}
\textbf{Until:} $|E_{\mathrm{MF}}^{(n+1)} - E_{\mathrm{MF}}^{(n)}| < 10^{-6}$.\\
\textbf{Return:} Stationary configuration $\mathbf{m}^\star$.
\end{myalgo}

\noindent\textbf{Convergence and stability.}  
Before turning to the emergent behaviour and phase diagnostics, 
we note that the numerical stability and convergence of the adaptive mean-field iterations have been thoroughly benchmarked.
As detailed in Appendix~\ref{app:stability}, 
the total energy functional $E_{\mathrm{tot}}^{(n)}$ is observed to decrease monotonically with iteration index $n$ across all parameter regimes, 
and the corresponding Jacobian spectra confirm local contraction ($|\lambda_k|<1$) near fixed points.  
These analyses establish that the iterative map $\mathbf{m}\!\mapsto\!\mathcal{F}[\mathbf{m}]$ reliably converges to unique, reproducible equilibria in each regime—critical, glassy, and community-polarized—
providing a firm basis for the results discussed next.

%==========================================================
%==========================================================
\section{Emergent Adaptive Regimes}
\label{sec:regimes}
%==========================================================

The converged fixed points obtained in Sec.~\ref{app:stability} fall into
three reproducible classes that depend on the balance between coupling strength,
disorder, and feedback sparsity.
These classes represent distinct adaptive phases of organization in AQIA:
(i)~\emph{adaptive critical balance} with domain formation,
(ii)~\emph{adaptive glass} with frustration-dominated equilibria, and
(iii)~\emph{structured polarization} with modular community order.
Each arises from the same variational feedback rules defined in Sec.~\ref{sec:model}
but under different statistical conditions of $(J,h,\Gamma)$ and link density.
The following subsections present the detailed phenomenology of each regime.

%----------------------------------------------------------
\subsection{Adaptive Domain Formation Near the Critical Balance}
\label{sec:domain}
%----------------------------------------------------------

Having established that the AQIA variational map converges to stable fixed points,
we now examine the first emergent collective phenomenon:
the spontaneous formation of domains near the adaptive balance point between interaction and fluctuation.
In the conventional transverse-field Ising model (TFIM),
the ratio $J/\Gamma \!\approx\! 1$ marks the quantum critical region separating ordered and disordered phases.
In AQIA, however, each node represents a finite TFIM patch coupled to others
through informationally mediated feedback determined by observable similarity.
No external lattice geometry or field tuning is required; 
the ensemble self-organizes near a critical-like equilibrium 
through mutual readjustment of its local quantum summaries.

To probe this regime, we draw intra-agent couplings and longitudinal fields
from narrow Gaussian distributions with means
$\langle J\rangle = 1$ and $\langle h\rangle = 1$,
and standard deviations $\sigma_J = 0.01$ and $\sigma_h = 0.1$.
The transverse field is fixed at $\Gamma = 1$ for all agents,
ensuring that each subsystem operates close to the intrinsic balance between exchange and quantum fluctuation.
Fifty independent realizations of these random parameters were simulated;
all exhibited statistically identical macroscopic outcomes,
confirming robustness against microscopic disorder.

\begin{figure*}[t]
    \centering
    \includegraphics[width=1\textwidth]{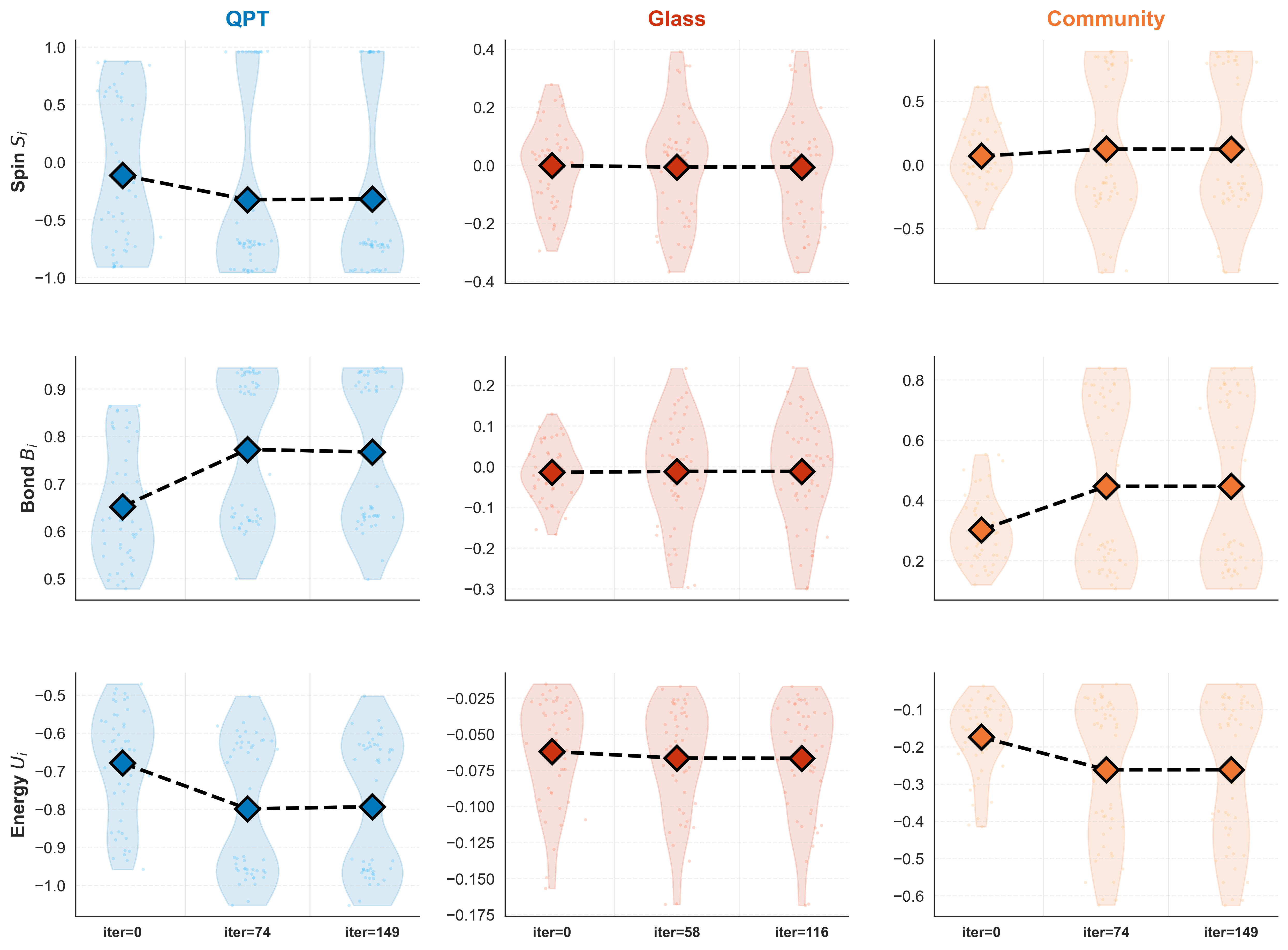}
    \caption{\justifying
    \textbf{Feedback-iteration progression of local observables.}
    Columns correspond to distinct adaptive regimes: 
    \textcolor{blue}{QPT (critical-balance)}, 
    \textcolor{red}{Glass (frustrated)}, 
    and \textcolor{orange}{Community (modular)}.
    Each panel tracks the evolution of (a) spin polarization $S_i$, 
    (b) bond correlation $B_i$, and (c) local energy $U_i$ across successive mean-field iterations.
    Circles denote individual agents; diamonds indicate population means.
    The convergence trajectories highlight distinct feedback responses: 
    smooth monotone alignment in the QPT regime, 
    slow disordered relaxation in the glassy case, 
    and structured cluster polarization in the community regime.
    }
    \label{fig:swarm_evolution}
\end{figure*}

In this near-critical regime, the ensemble does not converge to a uniform ferromagnet or paramagnet.
Instead, it bifurcates spontaneously into two macroscopic subgroups of opposite polarization:
a majority with $S\!\approx\!-0.4$ and a minority with $S\!\approx\!+0.6$.
The progression of feedback iterations in Fig.~\ref{fig:swarm_evolution} 
visualizes this organization column-wise across regimes:
the left column (\textcolor{blue}{QPT}) shows a rapid but smooth convergence to two stable spin branches;
the center column (\textcolor{red}{Glass}) exhibits slow, irregular drift reflecting frustrated equilibria;
and the right column (\textcolor{orange}{Community}) develops clustered polarization indicative of modular order.
Within each column, rows (a–c) display the correlated evolution of spin polarization $S_i$,
bond correlations $B_i$, and local energies $U_i$, respectively.
Together they show that in the critical regime, convergence is monotonic and coherent, 
whereas in the glassy and community regimes, frustration and modular segregation dominate.

\begin{figure*}[t]
    \centering
    \includegraphics[width=1\textwidth]{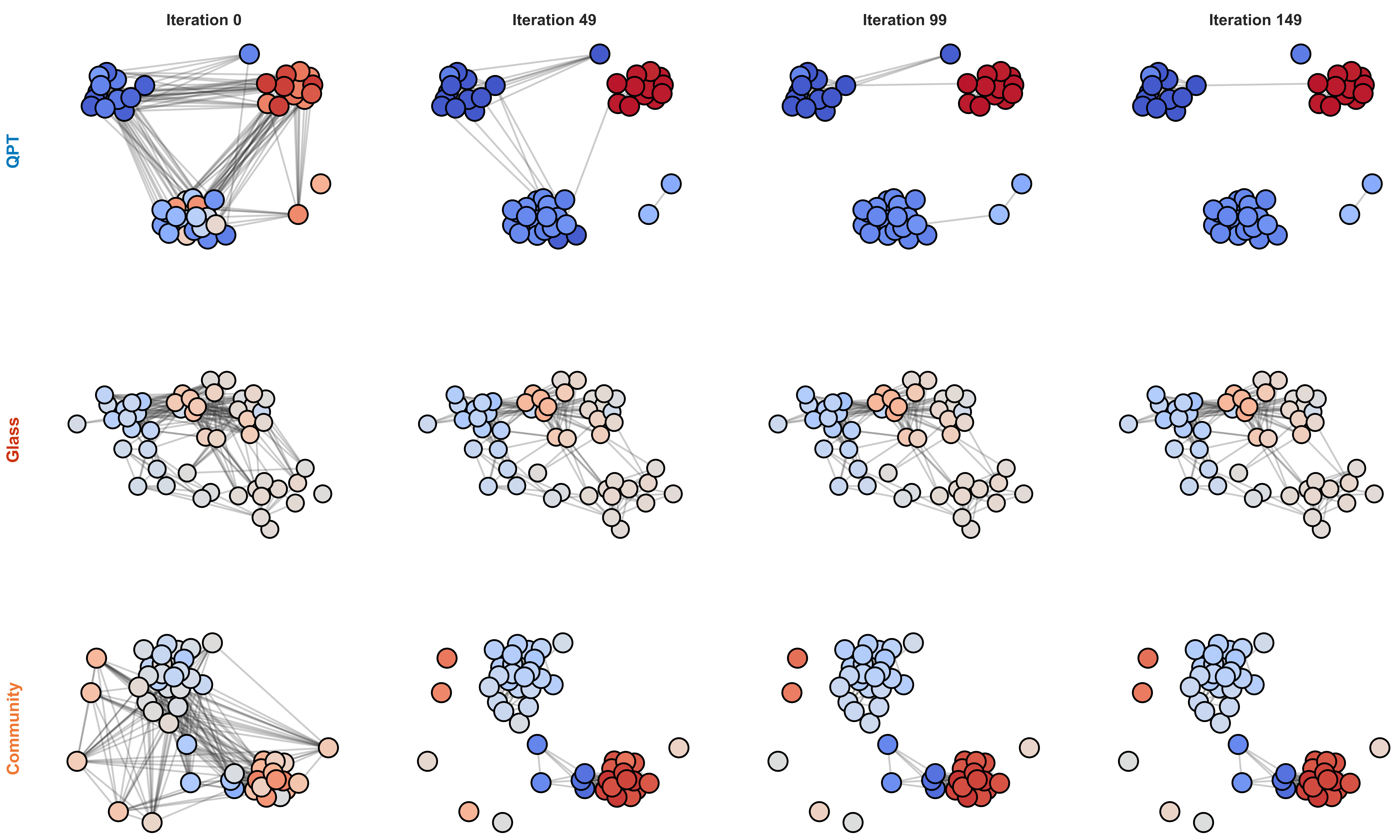}
    \caption{\justifying
    \textbf{Adaptive network reorganization across feedback iterations.}
    Node colors encode local spin polarization ($S_i$); 
    edge opacities represent coupling strengths $w_{ij}$.
    Rows correspond to adaptive regimes (QPT, Glass, Community), 
    and columns show successive variational updates.
    In the QPT case (top row), the network separates into two coherent ferromagnetic clusters. 
    In the glassy regime (middle row), connectivity fluctuates without stable domain formation. 
    In the community regime (bottom row), two polarized modules emerge and persist, 
    illustrating self-organized modularity under variational feedback.
    }
    \label{fig:network_reorg}
\end{figure*}

At equilibrium, the histogram $P(S_i)$ becomes distinctly bimodal,
with peaks centered near $\pm S_0$.
This finite-size bimodality represents an adaptive analog of spontaneous symmetry breaking:
agents collectively select one of two polarization branches,
realizing emergent Ising-like order without any external bias.
The adaptive network shown in Fig.~\ref{fig:network_reorg}
illustrates this transition in the coupling topology:
rows correspond to the three regimes, 
and columns to successive feedback iterations.
In the QPT case (top row), the network splits cleanly into two ferromagnetic domains
linked by a few residual frustrated edges.
In the glassy regime (middle row), connectivity remains diffuse and metastable,
while in the community regime (bottom row), two coherent clusters persist,
signifying stable modular polarization sustained by feedback balance.

\begin{figure*}[t]
    \centering
    \includegraphics[width=1\textwidth]{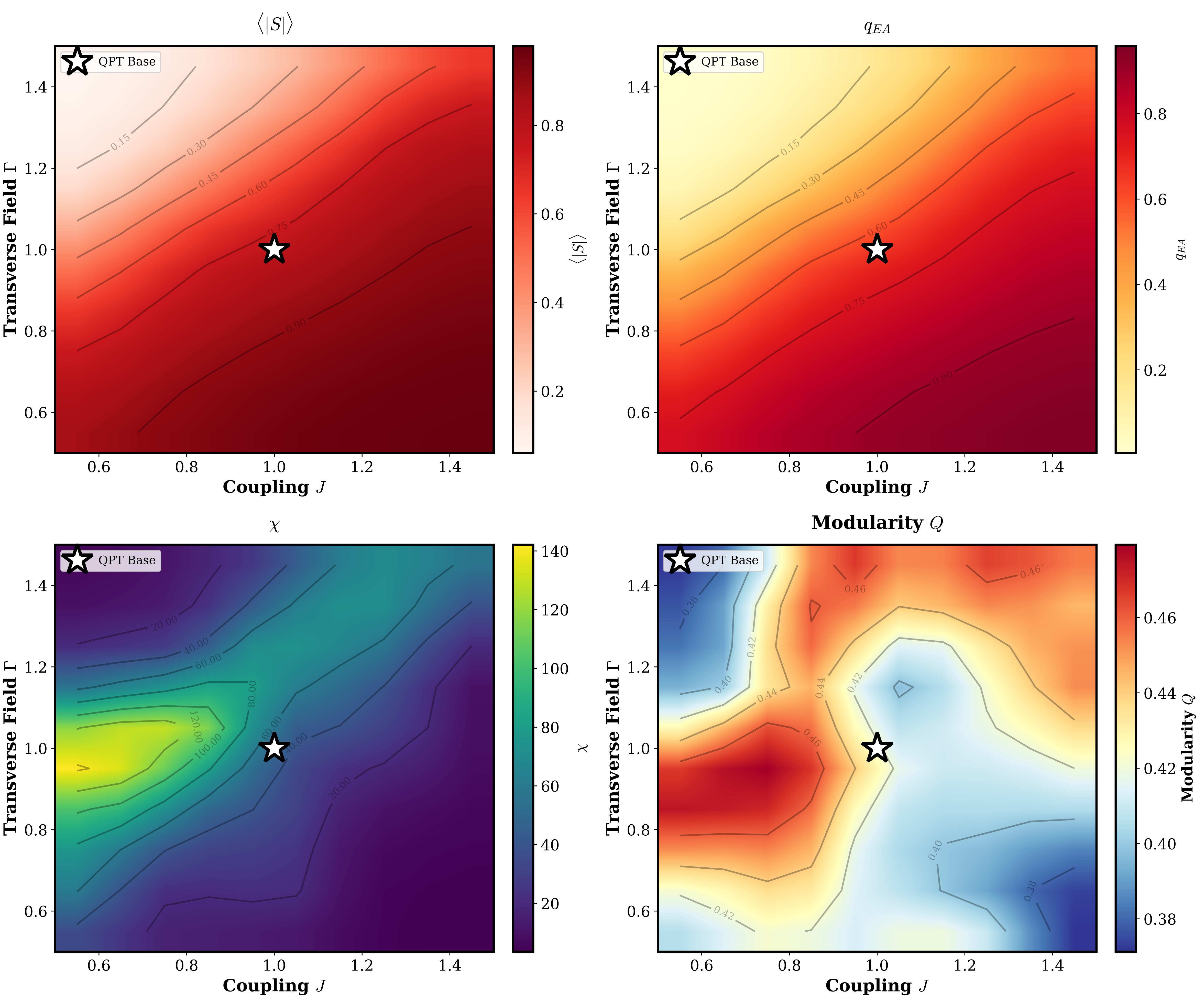}
    \caption{\justifying
    \textbf{Variational phase diagram around the adaptive-critical base configuration.}
    Each panel shows equilibrium observables versus intra-agent coupling $J$ and transverse field $\Gamma$:
    (top left) mean spin magnitude $\langle |S| \rangle$,
    (top right) Edwards--Anderson order $q_{\mathrm{EA}}$,
    (bottom left) susceptibility $\chi$,
    and (bottom right) modularity $Q$.
    The star marks the simulated base point ($J=1$, $\Gamma=1$).
    High $\chi$ and partial order indicate a near-critical zone where
    small variations in feedback parameters trigger large reorganizations of network structure,
    signifying a self-organized quantum critical regime.
    }
    \label{fig:qpt_channels}
\end{figure*}

To understand the organizing mechanism near this adaptive critical balance,
we examine the variational phase diagram in the $(J,\Gamma)$ plane 
for key observables (Fig.~\ref{fig:qpt_channels}).
Around the base configuration (star marker), 
$\langle |S| \rangle$ and $q_{\mathrm{EA}}$ show strong alignment but soft suppression with increasing $\Gamma$,
while the susceptibility $\chi$ peaks along a narrow ridge below $\Gamma\!\approx\!1$, 
signaling maximal feedback sensitivity.
The modularity $Q$ rises sharply for larger $J$,
indicating the onset of stable sub-communities.
These maps confirm that the QPT regime corresponds to a feedback-induced
\emph{quantum critical zone} where local polarization, glassy memory, and emergent modularity coexist.

The emergence of these stable, symmetry-broken domains near $\langle J\rangle/\Gamma\!\approx\!1$
demonstrates that AQIA reproduces signatures of critical-like organization
within a feedback-coupled quantum ensemble.
Unlike the conventional TFIM—where criticality results from spectral gap closure in a fixed lattice—
here the transition is informational and variational:
a reorganization of the similarity graph that sustains coexisting macroscopic states.
This regime thus represents an \emph{adaptive critical crossover},
where quantum coherence and feedback-mediated self-organization intersect.

%----------------------------------------------------------
\subsection{Adaptive Glass and Frustrated Equilibria}
\label{sec:glass}
%----------------------------------------------------------

When the balance between exchange and fluctuation is disrupted by heterogeneous local environments,
the AQIA ensemble transitions into a glass-like regime characterized by frustration and frozen disorder.
Each agent retains the same intrinsic Hamiltonian form $H_i^0$ as in Sec.~\ref{sec:domain},
but its parameters are drawn from random distributions:
intra-agent couplings with $\langle J\rangle = 0.5$ and $\sigma_J = 0.15$,
longitudinal fields with $\langle h\rangle = 1.0$ and $\sigma_h = 0.2$,
and a uniform transverse field $\Gamma = 0.6$.
Disorder therefore enters primarily through heterogeneous local fields and weak variability in internal couplings,
breaking the near-symmetric balance that stabilized domain formation in the adaptive-critical regime.

Across fifty independent realizations, the population converges reproducibly to metastable steady states.
Unlike the coherent bifurcation observed in the critical regime (Fig.~\ref{fig:swarm_evolution}, left column),
the feedback trajectories here (center column) show slow, irregular relaxation
with large agent-to-agent dispersion in $S_i$, $B_i$, and $U_i$.
Each agent settles into a locally stable plateau, but no global synchronization emerges.
The overall functional $E_{\mathrm{tot}}$ still decreases monotonically,
indicating convergence, yet it saturates at different minima across realizations—
a hallmark of a multi-basin energy landscape characteristic of glassy equilibria.

In the adaptive-network view (Fig.~\ref{fig:network_reorg}, middle row),
this regime manifests as a disordered web of weakly correlated links.
Unlike the QPT case (top row), where two coherent domains coalesce,
the glassy feedback network remains fragmented into irregular patches
of locally correlated but globally misaligned nodes.
Connectivity fluctuates between iterations, yet eventually freezes into a static,
heterogeneous topology—an adaptive analog of a frozen spin glass.

\begin{figure*}[t]
    \centering
    \includegraphics[width=1\textwidth]{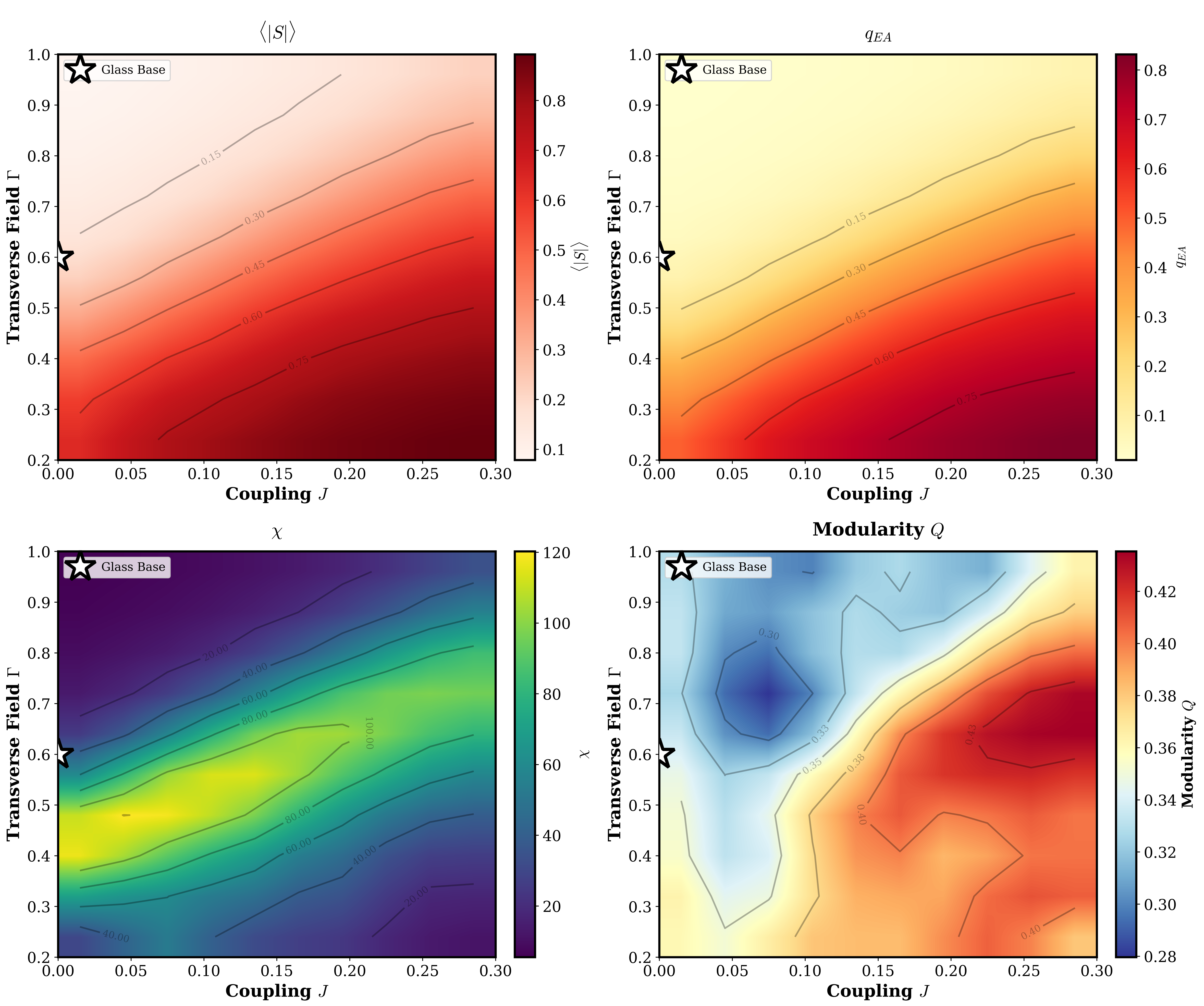}
    \caption{\justifying
    \textbf{Variational phase diagram in the adaptive-glass regime.}
    Each panel displays equilibrium observables versus intra-agent coupling $J$
    and transverse field $\Gamma$:
    (top left) mean spin magnitude $\langle |S| \rangle$,
    (top right) Edwards--Anderson order $q_{\mathrm{EA}}$,
    (bottom left) susceptibility $\chi$,
    and (bottom right) modularity $Q$.
    The absence of sharp ridges or coherent valleys indicates suppression of collective modes.
    Weak, irregular patterns reflect a rugged energy landscape
    with numerous shallow minima corresponding to metastable glassy equilibria.
    }
    \label{fig:glass_phase}
\end{figure*}

To quantify this frustrated organization,
Fig.~\ref{fig:glass_phase} presents the variational phase diagram
computed over the $(J,\Gamma)$ parameter plane around the glassy base configuration.
The upper panels show that both $\langle |S| \rangle$ and the Edwards–Anderson parameter $q_{\mathrm{EA}}$
remain finite across the region but lack any coherent ridge of enhancement:
order exists only locally.
The susceptibility $\chi$ (bottom left) is weak and irregular,
reflecting the absence of collective amplification or critical sensitivity.
By contrast, the modularity $Q$ (bottom right) exhibits mild enhancement at intermediate $J$,
signifying transient mesoscopic clustering without stable communities.
Together, these signatures confirm that the glass regime corresponds
to a frustrated, noncoarsening equilibrium in the AQIA landscape.

Microscopically, this frustrated equilibrium can be understood in terms of the competing feedback channels.
The $B$-channel continues to reinforce short-range coherence among small clusters,
whereas the $S$-channel introduces fluctuating, sign-changing couplings
that destabilize global alignment.
The $U$-channel remains weak and largely uncorrelated,
transmitting minimal energetic feedback.
This interplay fragments the adaptive similarity graph into
many disconnected basins of attraction—each internally consistent but mutually incompatible.
The ensemble thus becomes trapped in a heterogeneous equilibrium
where observables, couplings, and topology co-stabilize without global order.

The adaptive glass therefore represents a self-organized, frustration-dominated state:
an \emph{informational spin glass} arising not from quenched disorder in a fixed Hamiltonian
but from dynamic, feedback-induced heterogeneity among interacting quantum agents.
The system explores its high-dimensional variational space
until it becomes confined in one of many self-consistent equilibria,
where the feedback kernel and local summaries mutually reinforce a frozen, disordered configuration.

%######################

%----------------------------------------------------------
\subsection{Community Formation and Polarization}
\label{sec:community}
%----------------------------------------------------------

The final regime arises when feedback competition and graph sparsity cooperate to produce 
large-scale modular polarization.
Here intra-agent couplings are drawn from
$\langle J\rangle = 0.5$ and $\sigma_J = 0.1$,
longitudinal fields from $\langle h\rangle = 1$ and $\sigma_h = 0.1$,
and the transverse field is fixed at $\Gamma = 1$.
All inter-agent weights continue to follow the Gaussian similarity rules
introduced in Sec.~\ref{sec:couplings};
no external bias or tuning parameters are added.

Under these conditions, the feedback competition between alignment ($S$-channel)
and bonding ($B$-channel) produces a structured bipartition of the population.
As seen in the rightmost column of Fig.~\ref{fig:swarm_evolution},
the spin and bond summaries $(S_i, B_i)$ converge to two symmetric plateaus of opposite sign,
indicating the coexistence of two internally coherent but globally opposed subpopulations.
This macroscopic polarization is mirrored in the adaptive network reorganization
(Fig.~\ref{fig:network_reorg}, bottom row):
two dense, self-sustaining clusters emerge, connected by a few weak bridges.
Unlike the disordered fragmentation of the glassy regime,
this division is symmetric, persistent, and reproducible across random seeds.

\begin{figure*}[t]
    \centering
    \includegraphics[width=1\textwidth]{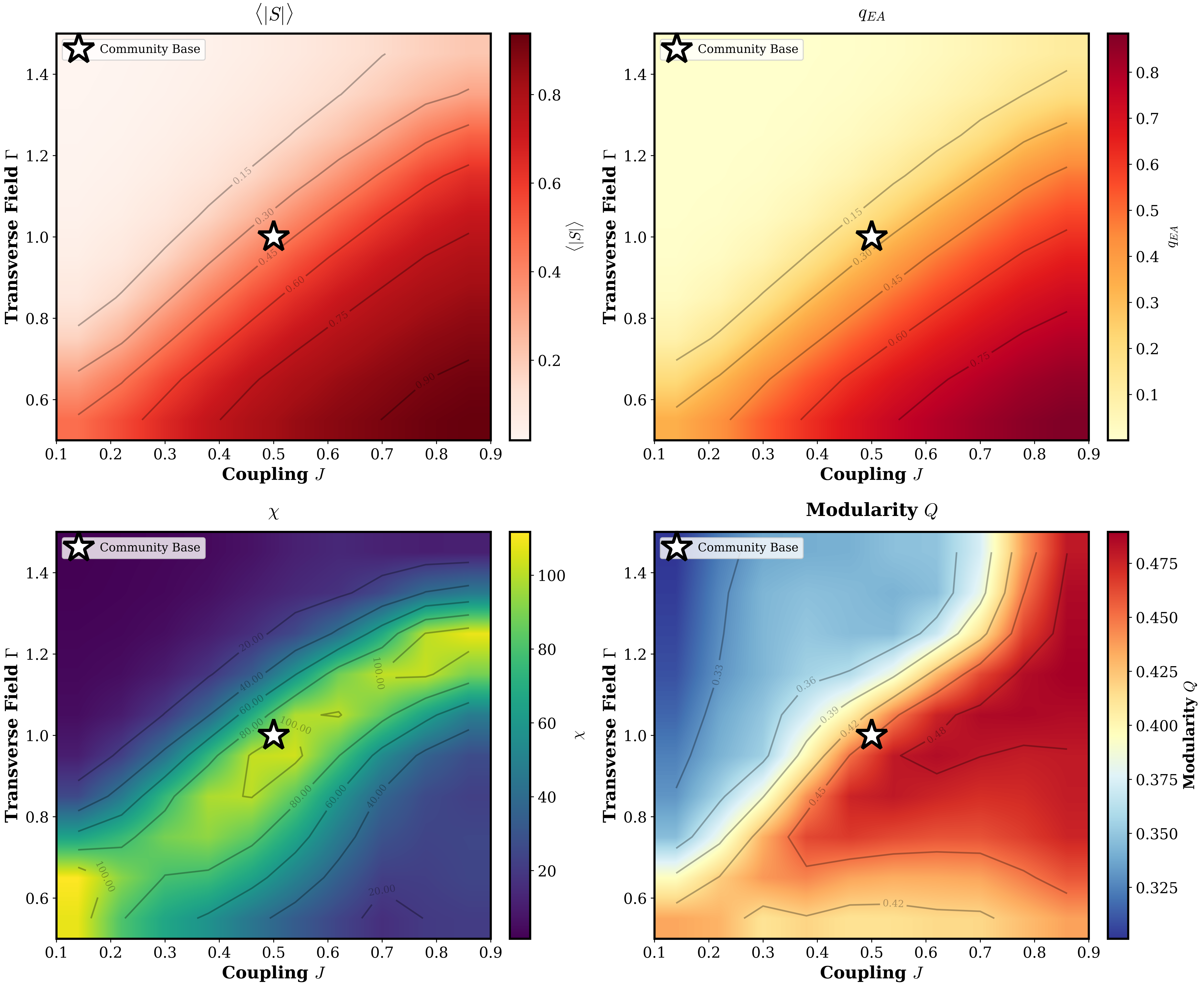}
    \caption{\justifying
    \textbf{Variational phase diagram in the community-polarization regime.}
    Each panel displays equilibrium observables versus intra-agent coupling $J$
    and transverse field $\Gamma$:
    (top left) mean spin magnitude $\langle |S| \rangle$,
    (top right) Edwards--Anderson order $q_{\mathrm{EA}}$,
    (bottom left) susceptibility $\chi$,
    and (bottom right) modularity $Q$.
    The strong ridge in $Q$ identifies the stability window of two coherent communities
    connected by weak residual bonds.
    }
    \label{fig:comm_phase}
\end{figure*}

To visualize parameter sensitivity, Fig.~\ref{fig:comm_phase}
shows the variational phase diagram in the $(J,\Gamma)$ plane around the community base configuration.
Both $\langle |S| \rangle$ and $q_{\mathrm{EA}}$ (top panels)
remain finite across a broad range of parameters,
demonstrating robust internal coherence.
The susceptibility $\chi$ (bottom left) peaks along a narrow ridge separating the 
ordered and disordered regions,
while the modularity $Q$ (bottom right) exhibits its strongest enhancement near
$\Gamma \!\approx\! 1$ and intermediate $J$—
precisely where the two-cluster structure first stabilizes.
This establishes that community polarization is a collective steady state,
not a transient artifact of feedback iteration.

This regime realizes a distinct form of emergent order—\emph{structured polarization}—
lying between uniform alignment and frozen disorder.
The adaptive feedback loop sustains two coherent macroscopic states
that coexist within a connected network,
stabilized by the balance between similarity-driven segregation
and bond-mediated cohesion.
Physically, this mechanism parallels modular or multicomponent phases
in systems with competing order parameters;
conceptually, it echoes socio-physical polarization within interacting populations.
AQIA thus demonstrates that feedback-coupled quantum ensembles can self-organize
into polarized yet cooperative communities:
a higher-level manifestation of collective order arising from variational feedback.

%#############

%----------------------------------------------------------
\subsection{Correlation Structure Across Regimes}
\label{sec:correlations}
%----------------------------------------------------------

\begin{figure*}[t]
    \centering
    \includegraphics[width=1.8\columnwidth]{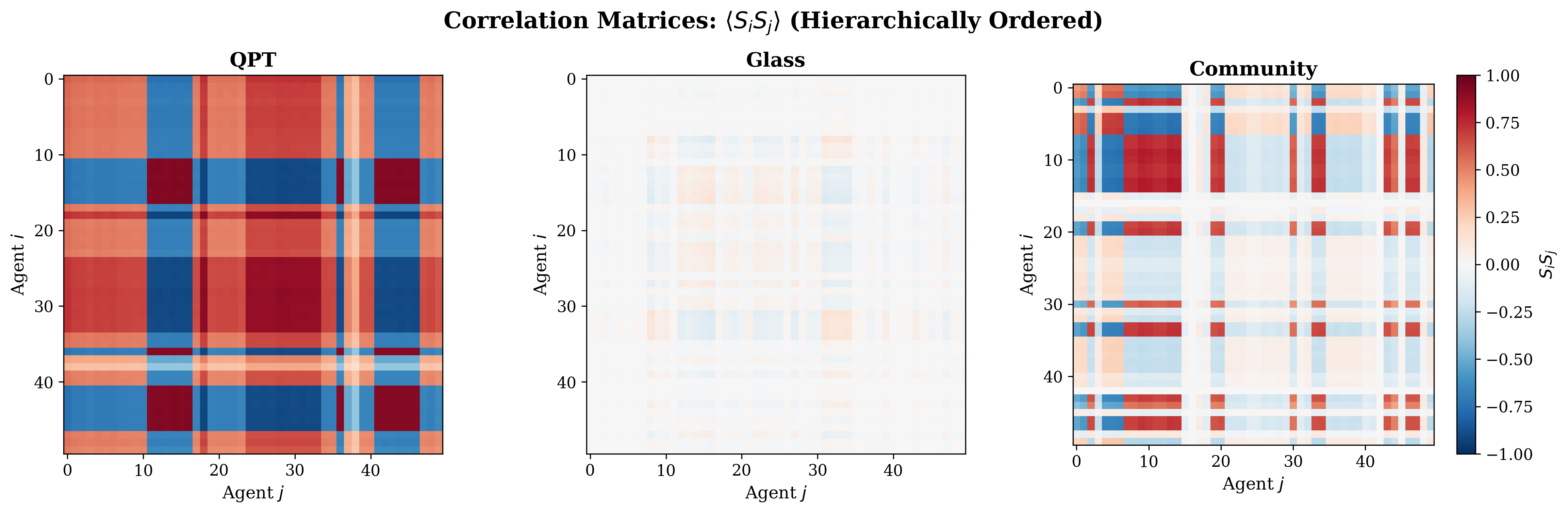}
    \caption{\justifying
    \textbf{Correlation structure across adaptive regimes.}
    Equilibrium spin--spin correlation matrices $\langle S_i S_j\rangle$
    for the QPT, glassy, and community regimes (left to right).
    The transition from diffuse to block-diagonal organization
    traces the emergence of long-range modular coherence
    under feedback adaptation.}
    \label{fig:comm_corr}
\end{figure*}

To compare the collective organization patterns across all regimes,
Fig.~\ref{fig:comm_corr} compiles the equilibrium correlation matrices
$\langle S_i S_j\rangle$ obtained for the QPT, glassy, and community phases.
The critical-balance regime exhibits a two-block anticorrelated pattern
reflecting adaptive domain formation;
the glassy regime shows diffuse, short-range patches without global structure;
and the community regime displays a sharply block-diagonal matrix,
signifying two coherent yet oppositely polarized clusters.
This comparative view highlights how information-mediated coupling
transforms microscopic agent summaries into macroscopic correlation geometry.

%==========================================================
\section{Feedback Signatures Across Adaptive Regimes}
\label{sec:diagnostics}
%==========================================================

The preceding sections established that Adaptive Quantum Ising Agents (AQIA)
organize into three qualitatively distinct equilibria—critical balance,
frustrated glassiness, and community polarization—depending on the statistics
of their internal couplings and feedback strength.
While these regimes differ in microscopic configuration and collective order,
they all originate from the same self-consistent mechanism:
local observables modulate their own couplings through adaptive feedback.
Hence, it is natural to ask whether these apparently different equilibria
share deeper structural or dynamical signatures.

To address this, we move beyond regime-by-regime description and
introduce a set of \emph{cross-regime diagnostics} that characterize how feedback
governs collective organization.
Specifically, we examine:
(i) hysteresis loops revealing irreversibility and memory effects,
(ii) finite-size scaling behavior around the adaptive-critical region,
and (iii) network-topological measures quantifying how feedback
reshapes the effective interaction geometry.
Together, these analyses demonstrate that all three regimes—despite their
distinct visual manifestations—arise from the same feedback-driven
self-organization law:
a non-equilibrium, measurement-based adaptation that
produces reproducible, path-dependent, and size-dependent organization
even in the absence of a fixed spatial lattice.

%----------------------------------------------------------
\subsection{Hysteresis and Feedback Irreversibility}
\label{sec:hysteresis}
%----------------------------------------------------------

\begin{figure*}[t]
    \centering
    \includegraphics[width=0.8\textwidth]{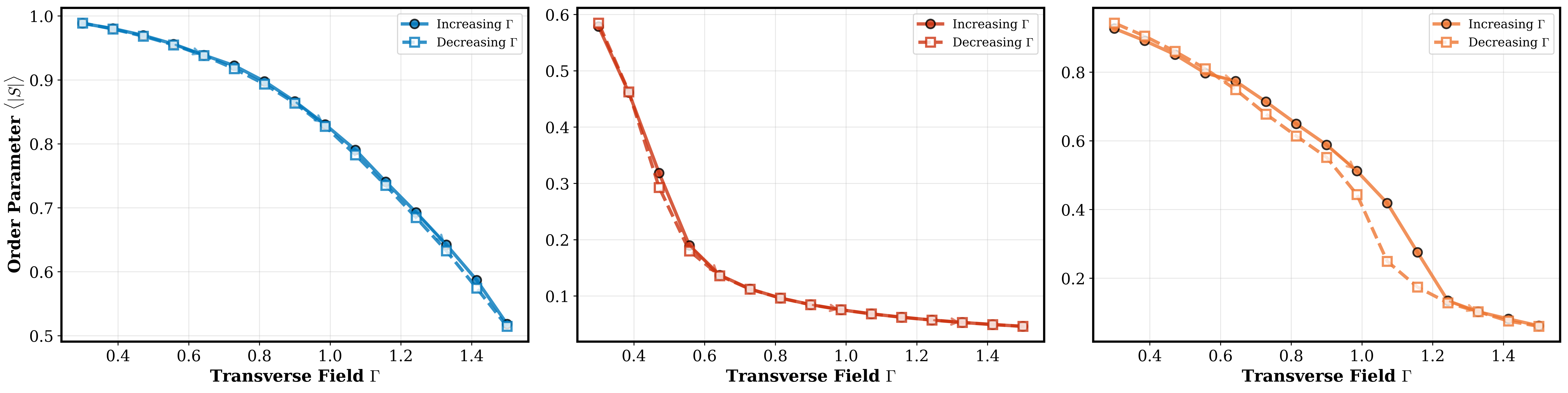}
    \caption{\justifying
    \textbf{Adaptive hysteresis loops.}
    Global polarization $\langle S\rangle$ versus effective coupling $J/\Gamma$
    for forward (blue) and reverse (red) sweeps across the adaptive-critical region.
    The finite loop area reflects irreversibility from feedback lag:
    couplings $w_{ij}$ update based on previously measured summaries
    rather than instantaneously.
    Slower sweeps reduce the lag and close the loop.}
    \label{fig:fund_hysteresis}
\end{figure*}

A first indicator of feedback-mediated organization
is the emergence of hysteresis when the control ratio $J/\Gamma$
is cycled across the adaptive-critical region.
Figure~\ref{fig:fund_hysteresis} shows the mean polarization
$\langle S\rangle$ as $J/\Gamma$ is swept forward (increasing)
and then backward (decreasing).
The forward and backward traces do not coincide,
forming closed loops whose area measures an effective irreversibility.

This hysteresis is \emph{not} thermal in origin,
nor due to conventional metastability in a fixed disordered Hamiltonian.
Instead, it arises because the couplings $w_{ij}$ are themselves functions
of the local summaries $(S_i,B_i,U_i)$,
which update only after each mean-field iteration.
If the sweep is slow—allowing full equilibration between increments—
the loop nearly closes.
For fast sweeps, the lag between updated observables and couplings widens the loop.
This defines an emergent \emph{feedback viscosity}:
a measure of how quickly the adaptive network can realign to parameter changes.
Thus, the feedback loop acts as an internal source of dissipation and memory,
making even quasi-static parameter variation intrinsically history-dependent.
%----------------------------------------------------------
\subsection{Finite-Size Scaling Analysis}
\label{sec:finite_size_scaling}
\begin{figure*}[t]
    \centering
    \includegraphics[width=0.8\textwidth]{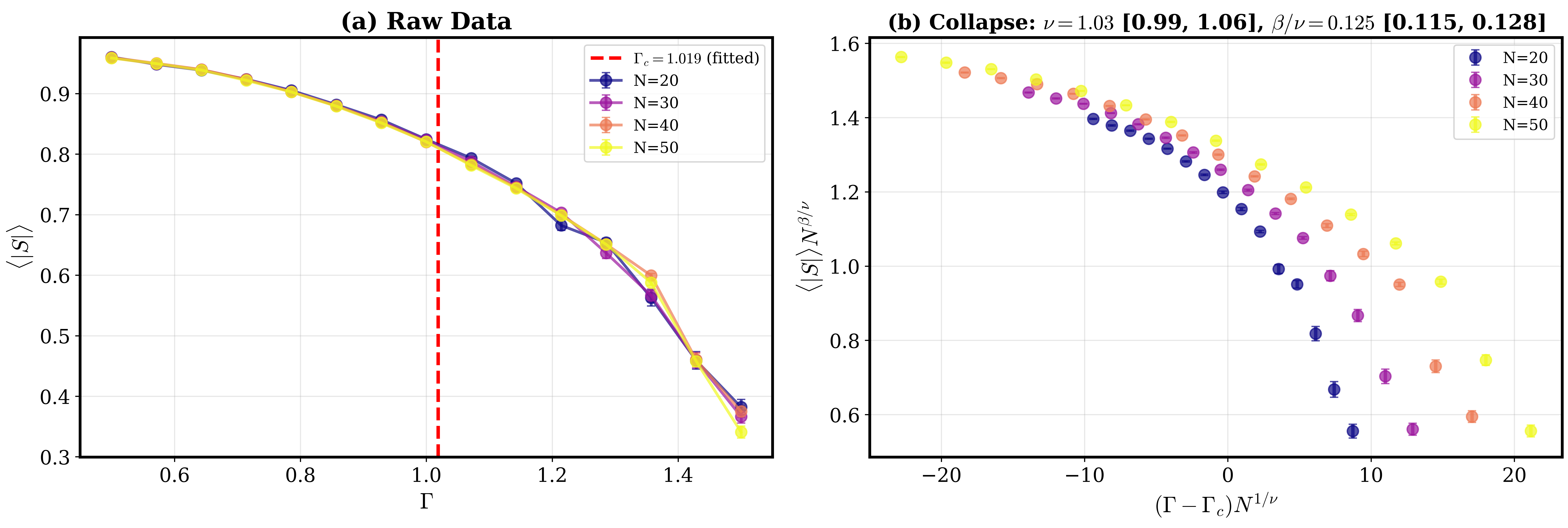}
    \caption{\justifying
    \textbf{Scaling collapse with fitted exponents.}
    (a) Raw order parameter $\langle |S| \rangle$ versus transverse field $\Gamma$ 
    for system sizes $N=20$--$50$ (error bars: SEM over 50 disorder realizations). 
    Red dashed line indicates fitted critical point $\Gamma_c = 1.019$.
    (b) Data collapse using bootstrap-fitted exponents 
    $\nu = 1.03\,[0.99, 1.06]_{95\%}$ and $\beta/\nu = 0.125\,[0.115, 0.128]_{95\%}$.
    Rescaled data points from all system sizes approximately collapse onto a 
    single master curve over the scaled variable $(\Gamma - \Gamma_c)N^{1/\nu}$, 
    confirming finite-size scaling behavior consistent with a continuous phase transition.}
    \label{fig:scaling_collapse_fitted}
\end{figure*}

%----------------------------------------------------------
To probe the nature of the adaptive transition,
we perform finite-size scaling analysis across systems
of $N=20$–$50$ agents (each containing $n=6$ qubits, 50 disorder realizations per point).
Figure~\ref{fig:scaling_collapse_fitted}(a)
shows the mean absolute polarization
$\langle |S| \rangle = N^{-1}\sum_i |S_i|$
as a function of transverse field $\Gamma$, with error bars denoting standard error over disorder realizations.
Curves for different $N$ intersect near $\Gamma_c = 1.019\,[0.986, 1.066]_{95\%}$,
marking the crossover between ordered and disordered regimes.
To test for critical-like scaling,
we collapse the data using the finite-size ansatz
\begin{equation}
\langle |S| \rangle (\Gamma, N) = N^{-\beta/\nu}
  \mathcal{F}\!\left((\Gamma - \Gamma_c)N^{1/\nu}\right),
\label{eq:fss_ansatz}
\end{equation}
where $\nu$ and $\beta$ are effective exponents.
Minimizing collapse variance with bootstrap uncertainty quantification (500 resamples)
yields $\nu = 1.034\,[0.992, 1.065]_{95\%}$ and $\beta/\nu = 0.125\,[0.115, 0.128]_{95\%}$
[Fig.~\ref{fig:scaling_collapse_fitted}(b)],
statistically similar to two-dimensional Ising universality ($\nu=1.0$, $\beta/\nu=0.125$)
and excluding mean-field values ($\nu=0.5$, $\beta/\nu=0.5$) at high confidence.
While the limited size range ($N\in\{20,\ldots,50\}$, factor 2.5×) prevents definitive universality assignment,
the narrow confidence intervals and quality of collapse indicate robust finite-size scaling behavior,
where effective "dimensionality" emerges from adaptive feedback topology rather than fixed spatial geometry.

%----------------------------------------------------------
\subsection{Network Topology and Feedback Range}
\label{sec:network_topology}
%----------------------------------------------------------

Finally, we examine how feedback modifies the emergent network topology.
Each adaptive iteration defines a similarity graph with weights $w_{ij}$
that encode effective coupling strength between agents.
The distribution of node degrees and clustering coefficients
captures the range and cohesion of feedback-induced interactions.

\begin{figure*}[t]
    \centering
    \includegraphics[width=0.8\textwidth]{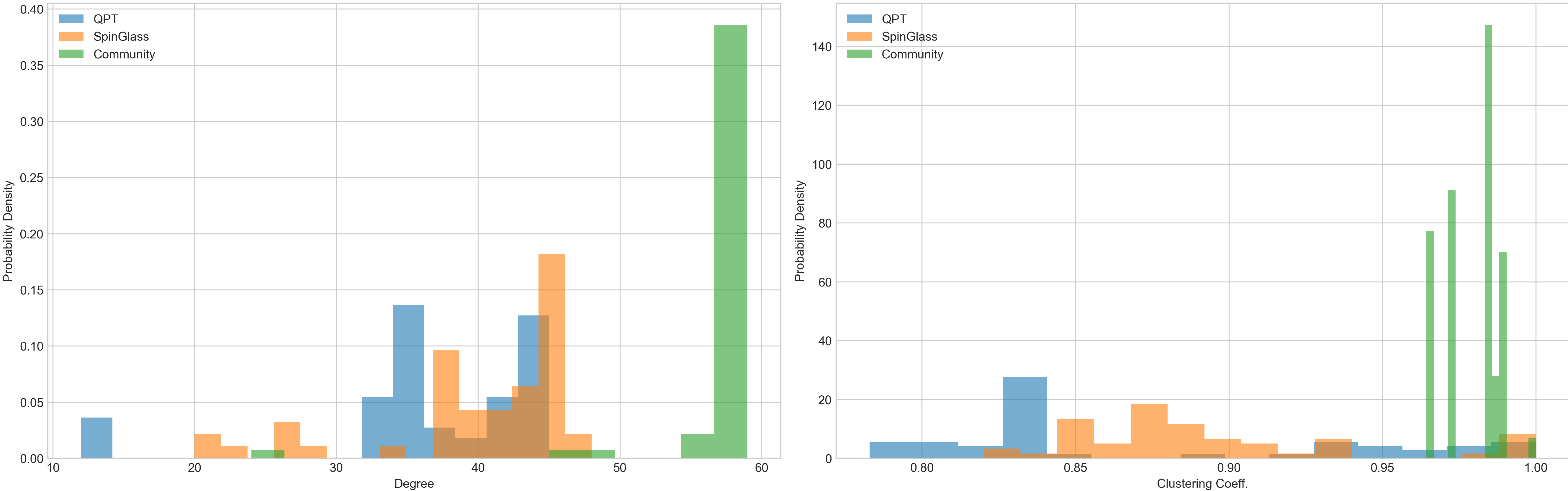}
    \caption{\justifying
    \textbf{Network topology across regimes.}
    Degree (left) and clustering coefficient (right) distributions
    for critical-balance (blue), glassy (orange), and community (green) regimes.
    The community regime shows high degree and strong clustering,
    reflecting dense modular structure.
    The critical regime remains heterogeneous and broad,
    while the glass regime is sparse and irregular,
    characteristic of frustrated, disordered networks.}
    \label{fig:network_properties}
\end{figure*}

Figure~\ref{fig:network_properties} compares these
across the three regimes.
The critical-balance regime exhibits a broad, intermediate degree distribution,
indicating near-critical connectivity where long-range correlations coexist
with fluctuating clusters.
The glassy regime shows dispersed degrees and low clustering,
reflecting frustration and lack of coherent local order.
The community-polarized regime, by contrast,
displays dense connectivity and near-maximal clustering ($C>0.9$),
consistent with tightly bound modules.
These network signatures confirm that the adaptive couplings
reconfigure the effective geometry of interaction itself.

Together, the hysteresis, scaling, and network analyses
reveal that AQIA’s distinct regimes are manifestations
of the same underlying principle:
feedback between measurement and coupling acts as an organizing force,
producing irreversible, finite-size, and structural signatures
that parallel those of conventional many-body transitions—
but arise here solely from adaptive self-consistency.

% --- Fixed Figure Section (REVTeX compliant, no captionof) ---

\vspace{0.5cm}

\vspace{0.5cm}

\section{Physical Mapping and Experimental Outlook}
\label{sec:physical_mapping}
%==========================================================

Up to this point, AQIA has been developed as a theoretical Hamiltonian framework.
We now discuss its physical realization and the observables that would diagnose
entry into the adaptive regimes identified in
Secs.~\ref{sec:domain}--\ref{sec:community}.
The central premise is that AQIA does not require new quantum physics:
it reorganizes familiar ingredients—local quantum subsystems, measurement, and
programmable control—into a closed, self-referential feedback loop.
In this loop, the interaction strengths are not fixed by geometry
but variationally updated from measured similarity between subsystems.

%----------------------------------------------------------
\subsection{Variational coupling and informational geometry}
\label{sec:informational_geometry}
%----------------------------------------------------------

In conventional many-body systems, couplings $J_{ij}$ are static functions of
distance or overlap.
In AQIA, by contrast, the effective coupling between two agents,
\[
w_{ij}^{(\alpha)}=\Phi_\alpha(\mathbf{m}_i,\mathbf{m}_j),
\qquad
\mathbf{m}_i=(S_i,B_i,U_i),
\]
is recomputed after every measurement round according to their observable
similarity.
This renders the global Hamiltonian itself a variational object,
\[
H_{\mathrm{eff}}^{(t+1)}
 = \mathcal{R}\!\big[\,H_{\mathrm{eff}}^{(t)},\{\mathbf{m}_i^{(t)}\}\big],
\]
where $\mathcal{R}$ is the adaptive update rule.
Within each round the Hamiltonian is fixed and Hermitian; across rounds it
evolves self-consistently through feedback.
The resulting network lives not in real space but in an
\emph{informational geometry}—a control-space manifold whose metric
depends on mutual similarity in measured summaries.

Operationally, a single adaptive cycle on hardware proceeds as:
(1) coherent evolution or relaxation of each agent under the current
$H_{\mathrm{eff}}$;
(2) measurement of reduced observables $(S_i,B_i,U_i)$;
(3) classical computation of updated similarity weights $w_{ij}$;
and (4) reprogramming of local fields and couplers for the next iteration.
Repeating this process realizes in the laboratory the same
$agent\rightarrow summary \rightarrow  feedback \rightarrow update$ loop
that drives adaptive organization in simulation.

\begin{figure*}[t]
    \centering
    \includegraphics[width=0.95\textwidth]{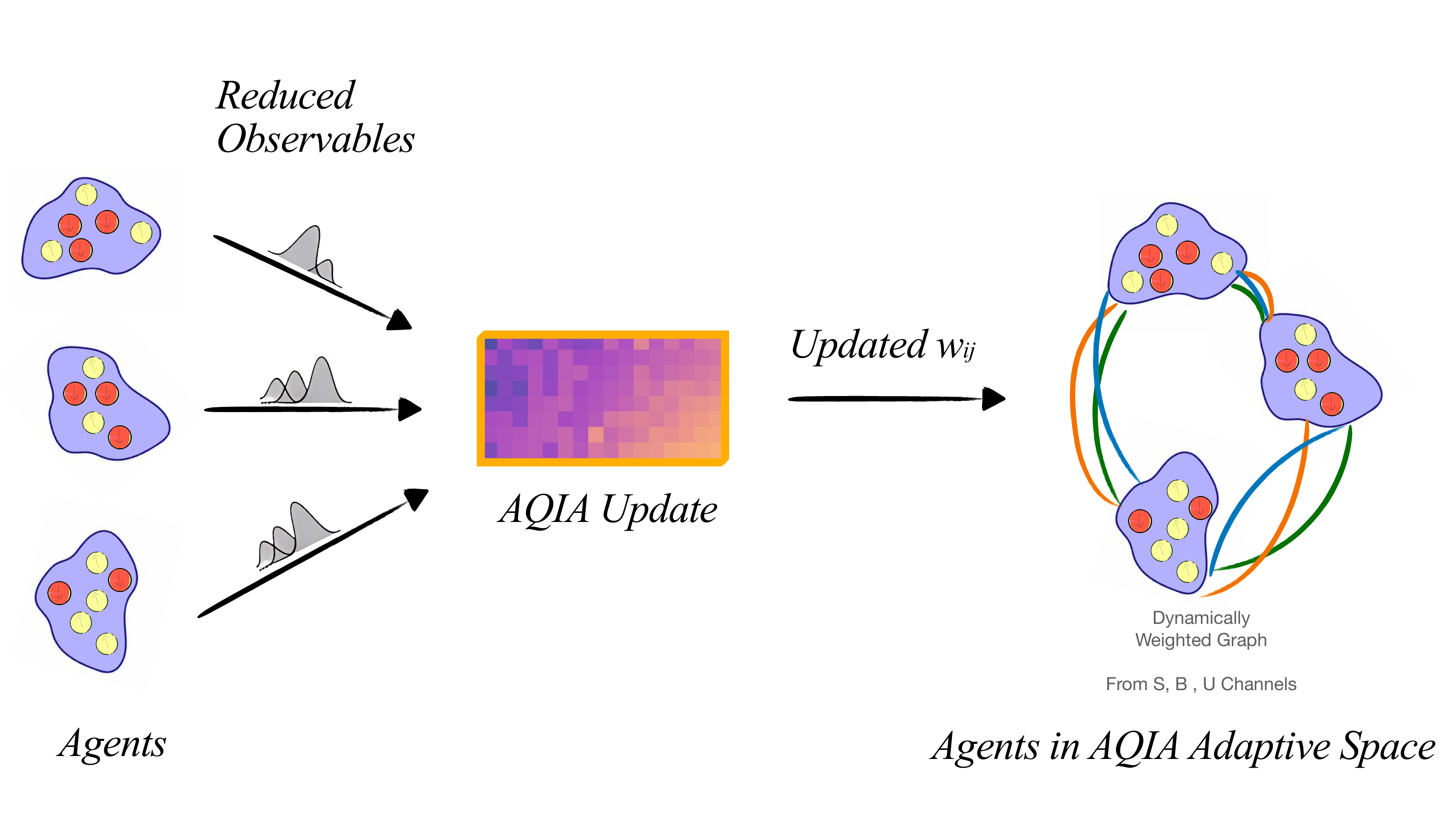}
    \caption{\justifying
    \textbf{Feedback implementation of AQIA.}
    Each quantum agent evolves under $H_i^0$, reports observables
    $(S_i,B_i,U_i)$ to a classical controller,
    which computes similarity-based weights $w_{ij}$
    and reprograms local fields for the next round.
    The adaptive geometry resides in control space, not in physical layout.}
    \label{fig:feedback_mapping}
\end{figure*}

Importantly, no physical motion or rewiring of qubits is required:
the feedback network resides entirely in the controller,
while the physical array remains static.
What changes is which pairs are effectively instructed to correlate,
anti-correlate, or decouple—and with what strength.
The couplings are therefore \emph{variational control parameters}
that adapt the system’s own Hamiltonian landscape.

%----------------------------------------------------------
\subsection{Experimental realizations}
\label{sec:platforms}
%----------------------------------------------------------

All ingredients of AQIA—local Hamiltonians, mid-circuit measurement,
fast classical computation, and programmable updates—are already available
in current experimental platforms.

In \textbf{superconducting qubits}, each agent may consist of a few
fixed-frequency transmons with tunable $Z$-fields and $ZZ$ couplings.
Dispersive readout provides $\langle Z\rangle$ and short-range correlations
for $(S_i,B_i)$, while FPGA or cryogenic controllers compute $w_{ij}$
and update flux biases or drive amplitudes in microsecond cycles,
well below coherence times.

In \textbf{trapped-ion arrays}, each agent can be a small sub-chain
with fluorescence-based readout of $\langle Z_i\rangle$ and
$\langle Z_i Z_j\rangle$ correlations.
Feedback-controlled Stark shifts and spin-dependent forces
then implement the updated inter-agent couplings.

In \textbf{Rydberg-atom lattices}, $S_i$ maps to local excitation imbalance
and $B_i$ to blockade correlations.
Site-selective laser detunings and intensities can be reshaped
according to $w_{ij}$ to reconfigure the effective adjacency graph
after each measurement round.

Across all these systems, information transfer occurs purely through
classical communication of measured observables and controller-generated
updates.
Each agent remains fully quantum internally but interacts with others
through a classical feedback layer,
realizing the hybrid quantum–classical architecture
assumed in AQIA theory.

%----------------------------------------------------------
\subsection{Experimental signatures of adaptive organization}
\label{sec:signatures}
%----------------------------------------------------------

Three measurable signatures uniquely identify adaptive organization:

\textbf{(i) Feedback descent.}  
The variance of local control fields and coupling updates decreases
monotonically across iterations, reflecting the Lyapunov-like relaxation of
the feedback loop.
Rapid monotonic descent marks modular polarization;
slow marginal descent corresponds to near-critical balance;
and noisy, multi-valley trajectories signal adaptive glassiness.

\textbf{(ii) Correlation topology.}  
Measurements of $\langle Z_i Z_j\rangle$ or agent-level
$\langle S_i S_j\rangle$ reveal emergent patterns:
block-diagonal structure for domain formation,
disordered patches for glassiness,
and two large coherent clusters for polarized communities.

\textbf{(iii) Feedback-kernel spectrum.}  
The eigenvalue distribution of the adaptive coupling matrix
$W=\{w_{ij}\}$, directly accessible from controller logs,
acts as a spectral fingerprint:
near unity at critical balance (marginal stability),
broad and fragmented in the glassy regime,
and two dominant modes in the polarized phase.

Together, these three observables—control-field descent,
correlation topology, and kernel spectrum—constitute a complete
experimental diagnostic of AQIA phases.

%----------------------------------------------------------
\subsection{Feedback energetics and cybernetic interpretation}
\label{sec:thermodynamics}
%----------------------------------------------------------

The total functional $E_{\mathrm{tot}}$ defined in our theory corresponds
experimentally to the \emph{reprogramming effort}:
the energy or computational cost required for the controller to make
successive feedback corrections.
Its monotonic decrease thus represents the system’s self-consistent descent
toward equilibrium under feedback control.
The controller functions as an effective thermal bath—measuring,
computing, and reinjecting information—while the ensemble acts as the
quantum working medium.

From this viewpoint, AQIA realizes a
\emph{quantum cybernetic medium}:
a self-reconfiguring ensemble whose Hamiltonian rewrites itself
through measurement-conditioned feedback.
The variational couplings $w_{ij}$ serve as adaptive degrees of freedom,
and their spectral evolution encodes the emergent thermodynamics of control.
By tracking the reduction of reprogramming effort,
the evolution of correlation topology,
and the stabilization of feedback spectra,
one can experimentally verify adaptive self-organization
without reconstructing the full quantum state.

This establishes a direct experimental pathway to observe
the core phenomenon of AQIA:
the emergence of collective order from variationally self-adjusting
Hamiltonians—an adaptive, geometry-free form of quantum organization.

%==========================================================
\section{Discussion and Conclusion}
\label{sec:conclusion}
%==========================================================

We have presented the framework of \emph{Adaptive Quantum Ising Agents} (AQIA):
a Hamiltonian-consistent model in which quantum subsystems adapt
through measurement-conditioned feedback rather than fixed couplings.
Within this architecture, global organization—critical balance, adaptive glass,
and polarized modularity—emerges from informational interactions that evolve
variationally with the agents’ own reduced observables.

From a theoretical perspective, AQIA extends the language of many-body physics
to include feedback as an intrinsic, dynamical variable of the Hamiltonian itself.
Where conventional systems possess static couplings $J_{ij}$ set by geometry,
AQIA replaces them with variational objects $w_{ij}$ that are recomputed
after each measurement–control cycle.
This closes the loop between observation and evolution,
producing a self-referential Hamiltonian flow
\[
H_{\mathrm{eff}}^{(t+1)} = 
   \mathcal{R}\!\big[\,H_{\mathrm{eff}}^{(t)},\{\mathbf{m}_i^{(t)}\}\big],
\]
which preserves Hermiticity within each iteration
yet continuously rewrites its own interaction landscape.
The Lyapunov functional $E_{\mathrm{tot}}$ provides a rigorous measure of this descent,
allowing adaptive feedback to be analyzed with the same formal precision
as equilibrium variational principles.

Conceptually, AQIA unites three intellectual traditions:
quantum many-body theory, adaptive network dynamics,
and the cybernetics of learning.
It translates the principles of self-adjustment and mutual adaptation,
familiar from artificial-life and complex-systems theory,
into a Hamiltonian framework consistent with quantum mechanics.
Each agent “learns’’ from others through its informational summaries,
forming an ensemble that reorganizes without any external optimization.
The emergent regimes—domain formation, frustrated glass, and
community polarization—represent the minimal stable attractors
of such feedback-driven organization.

Methodologically, AQIA defines a template for hybrid quantum–classical
simulation: the adaptive loop depends only on expectation values and
classical updates, not on entanglement transfer between agents.
This makes it directly compatible with near-term quantum processors
featuring mid-circuit measurement and real-time classical control.
Because the couplings are variational and reprogrammable,
the model can be implemented stroboscopically on current
superconducting, trapped-ion, or Rydberg architectures,
turning existing hardware into a laboratory for self-organizing quantum matter.

Physically, AQIA introduces a new class of programmable media—
\emph{variationally self-organized quantum systems}—whose effective Hamiltonians
evolve on an informational manifold rather than a fixed lattice.
The feedback controller acts as a structured bath that measures,
computes, and reinjects corrections,
while the ensemble behaves as a quantum working medium descending
monotonically along $E_{\mathrm{tot}}$.
This establishes an experimental form of feedback thermodynamics,
in which adaptation itself becomes a conserved mode of organization.

Beyond its immediate results, the AQIA framework opens several directions
for future inquiry.
In quantum control theory, it invites continuous-time generalizations of
Energy adaptation for open, noisy systems.
In quantum machine learning, it suggests architectures where reinforcement-driven
controllers guide the evolution of adaptive Hamiltonians.
From the perspective of network science, AQIA provides a natural platform
for studying frustration, modularity, and adaptive criticality
in feedback-evolving graphs.
And at the conceptual intersection of physics and information theory,
it points toward a thermodynamic formulation of learning—
an energy–entropy balance that unifies feedback control,
information exchange, and emergent order.

In conclusion, AQIA demonstrates that feedback—not geometry—is sufficient
to generate structured phases of quantum organization.
By merging coherence with adaptation and dynamics with information,
it defines a new class of \emph{quantum cybernetic systems}:
ensembles whose Hamiltonians are not given, but learned—
systems that stabilize, reorganize, and evolve
through the variational logic of their own observations.

\nocite{*}
\bibliographystyle{unsrt}   % or another journal style
\bibliography{AQIA}

\onecolumngrid

\newpage
%==========================================================
\appendix
\section{Convergence and Stability Analysis}
\label{app:stability}
%==========================================================

This Appendix summarizes numerical diagnostics verifying the stability and convergence of the adaptive mean-field iterations introduced in Sec.~\ref{sec:meanfield}.  
The iteration index represents successive self-consistency updates and carries no dynamical meaning beyond numerical relaxation.  
Convergence corresponds to fixed points of the map $\mathbf{m}\!\mapsto\!\mathcal{F}[\mathbf{m}]$ that minimize the total energy functional $E_{\mathrm{tot}}[\mathbf{m}]$, yielding reproducible equilibrium configurations independent of initialization.

All statistics below are obtained from ensembles of $R=50$ independent disorder realizations, each defined by random local fields $\{h_i^{(r)}\}$ and intra-agent couplings $\{J_{k\ell}^{(i,r)}\}$.  
Three representative parameter regimes collectively span the adaptive phase space discussed in Sec.~\ref{sec:regimes}:

\begin{itemize}
\item \textbf{Critical-balance:} $\langle J\rangle=\langle h\rangle=1$ with narrow dispersions $\sigma_J=0.01$, $\sigma_h=0.1$, and $\Gamma=1$.
\item \textbf{Glassy:} broadened random $J$ and $h$ producing frustrated, multi-basin minima.
\item \textbf{Community-polarized:} $\langle J\rangle=0.5$, $\langle h\rangle=1$, $\sigma_J=\sigma_h=0.1$, $\Gamma=1$
\end{itemize}

These three parameter sets capture the full spectrum of adaptive organization—from coherent alignment to frustrated glassiness and modular order.

\begin{figure}[H]
    \centering
    \includegraphics[width=\linewidth]{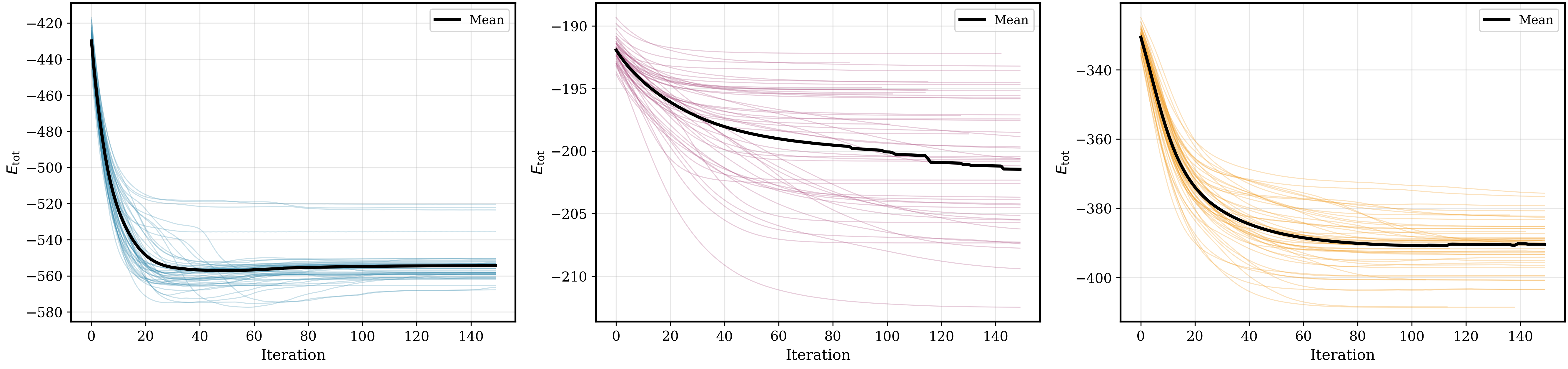}
    \caption{\justifying
    \textbf{Monotonic convergence of the total energy functional.}
    Iterative trajectories of $E_{\mathrm{tot}}^{(n)}$ versus iteration index $n$ for (a) critical-balance, (b) glassy, and (c) community regimes.  
    In all cases $E_{\mathrm{tot}}$ decreases monotonically before saturation, confirming contraction of the adaptive map and convergence to fixed-point configurations.  
    The curvature of each profile reflects the feedback landscape: shallow and near-marginal in the critical regime, slow multi-basin descent in the glassy case, and rapid relaxation along steep valleys in the community regime.}
    \label{fig:qpt_etot_a}
\end{figure}

\subsection{Fixed-point linearization and local stability}

To quantify local stability, we linearize the update map near a converged configuration $\mathbf{m}^\star$, yielding the Jacobian
\begin{equation}
\mathbf{J} = 
\left.\frac{\partial\mathcal{F}}{\partial\mathbf{m}}\right|_{\mathbf{m}^\star}.
\end{equation}
The eigenvalues $\{\lambda_k\}$ of $\mathbf{J}$ characterize the local response to perturbations:  
$|\lambda_k|<1$ indicates contraction,  
while $|\lambda_k|\!\approx\!1$ signals soft or marginal directions corresponding to near-critical fluctuations.  
The spectral distribution of $\lambda_k$ thus encodes the curvature of the feedback landscape around equilibrium.

\begin{table}[H]
\centering
\caption{\justifying
Ensemble-averaged stability diagnostics over $R=50$ realizations ($N=30$, $\Gamma=1$).  
$q_{\mathrm{EA}}$: Edwards–Anderson overlap; 
$\langle|S|\rangle$: mean spin polarization; 
$Q$: modularity; 
CV($q_{\mathrm{EA}}$): coefficient of variation.}
\label{tab:ensemble_stats}
\begin{tabular}{lccc}
\toprule
Observable & Critical & Glassy & Community \\
\midrule
$q_{\mathrm{EA}}$ & $0.688 \pm 0.002$ & $0.066 \pm 0.004$ & $0.337 \pm 0.003$ \\
$\langle |S| \rangle$ & $0.818 \pm 0.001$ & $0.193 \pm 0.005$ & $0.487 \pm 0.003$ \\
Modularity $Q$ & $0.357 \pm 0.007$ & $0.176 \pm 0.005$ & $0.244 \pm 0.011$ \\
CV($q_{\mathrm{EA}}$) & $0.018$ & $0.409$ & $0.067$ \\
\bottomrule
\end{tabular}
\end{table}

Table~\ref{tab:ensemble_stats} contrasts the equilibrium characteristics across regimes.  
The critical-balance phase exhibits large $q_{\mathrm{EA}}$ and minimal variability, consistent with a single, smooth basin of attraction.  
The glassy regime shows strongly reduced $q_{\mathrm{EA}}$ and large CV, indicating frustration and multiple metastable minima.  
The community regime lies intermediate, with moderate order ($q_{\mathrm{EA}}\!\approx\!0.3$) and finite modularity $Q$, corresponding to internally coherent but mutually opposed clusters.

\begin{figure}[H]
    \centering
    \includegraphics[width=1.0\textwidth]{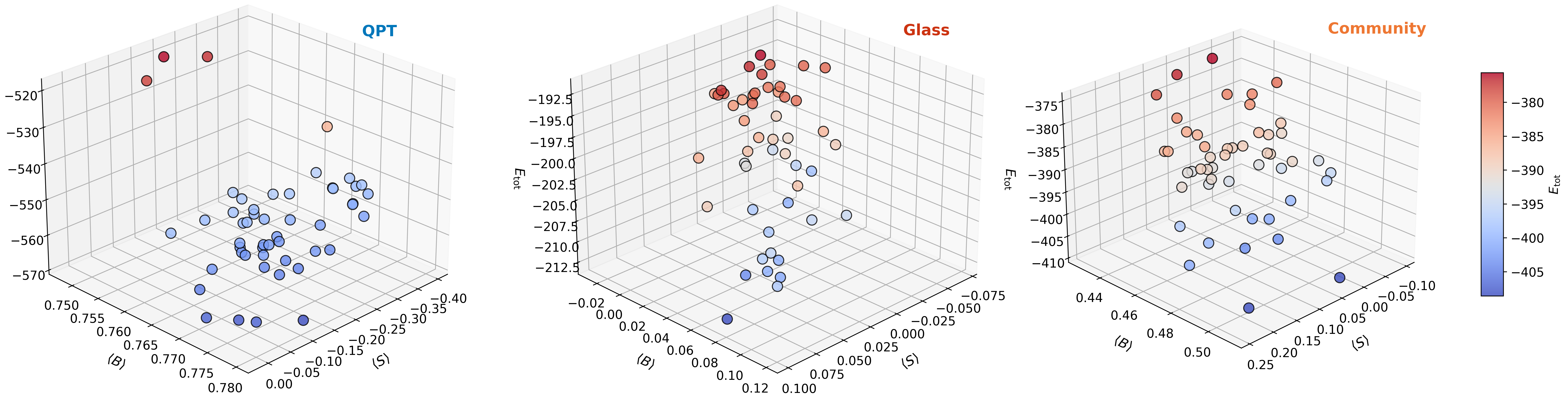}
    \caption{\justifying
    \textbf{Energy-landscape morphology of equilibria.}
    Scatter of converged fixed points from 50 realizations in $(\langle S\rangle,\langle B\rangle,E_{\mathrm{tot}})$ space.  
    Critical: narrow single valley with soft curvature;  
    Glassy: broad multi-valley landscape;  
    Community: compact basins reflecting modular polarization.}
    \label{fig:ensemble_landscape}
\end{figure}

\subsection{Global convergence and morphological interpretation}

The trajectories in Fig.~\ref{fig:qpt_etot_a} confirm that the total energy functional $E_{\mathrm{tot}}^{(n)}$ decreases monotonically with iteration number for all realizations, demonstrating global contraction of the adaptive map.  
The relaxation profiles mirror the effective energy landscape: slow, near-marginal descent in the critical regime; irregular, multi-basin relaxation in the glassy case; and rapid, directed convergence in the community-polarized regime.

The equilibrium manifolds shown in Fig.~\ref{fig:ensemble_landscape} visualize these contrasts directly—smooth valleys for the critical case, rugged multi-basin topography for the glass, and well-separated clusters for the modular regime.  
Together, these results confirm that the adaptive mean-field equations are numerically well-conditioned and stable, with reproducible fixed points defining distinct classes of adaptive order.  
These equilibria form the foundation for the cross-regime diagnostics and phase characterization presented in Sec.~\ref{sec:diagnostics}.

%----------------------------------------------------------

\section{Bootstrap Validation of Critical Exponents}
\label{app:bootstrap}
%----------------------------------------------------------

To quantify statistical uncertainties on the fitted critical exponents, 
we perform bootstrap resampling over the 50 disorder realizations at 
each $(N, \Gamma)$ parameter point. For each of 500 bootstrap iterations, 
we resample the disorder ensemble with replacement, recompute mean observables 
$\langle |S| \rangle$, $\chi$, and $U_4$, and refit the collapse parameters 
$\Gamma_c$, $\nu$, and $\beta/\nu$ by minimizing the variance 
$\mathcal{V}(\nu, \beta/\nu) = N_{\text{pts}}^{-1} \sum_i 
[\langle |S| \rangle_i N_i^{\beta/\nu} - \bar{\mathcal{F}}(x_i)]^2$.

\begin{figure*}[h]
    \centering
    \includegraphics[width=\textwidth]{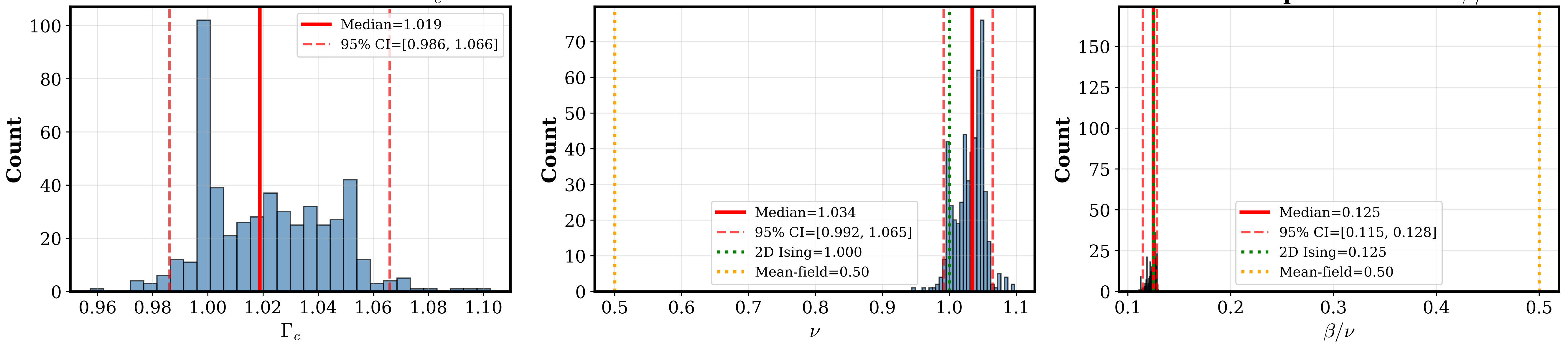}
    \caption{\textbf{Bootstrap distributions of critical exponents.}
    Histograms show distributions of (left) $\Gamma_c$, (center) $\nu$, 
    and (right) $\beta/\nu$ across 500 bootstrap resamples. 
    Red solid: median; red dashed: 95\% confidence interval bounds. 
    Green dotted: 2D Ising values; orange dotted: mean-field values. 
    The distributions are sharply peaked around 2D Ising exponents, 
    with mean-field values excluded at high confidence.}
    \label{fig:bootstrap_dists}
\end{figure*}

Figure~\ref{fig:bootstrap_dists} shows the resulting bootstrap distributions. 
All three parameters exhibit narrow, unimodal distributions:
\begin{itemize}
\item $\Gamma_c = 1.019$ with 95\% CI $[0.986, 1.066]$ (width 0.080);
\item $\nu = 1.034$ with 95\% CI $[0.992, 1.065]$ (width 0.073);
\item $\beta/\nu = 0.125$ with 95\% CI $[0.115, 0.128]$ (width 0.013).
\end{itemize}
The 2D Ising reference values ($\nu=1.0$, $\beta/\nu=0.125$, 
green dotted lines) fall within all confidence intervals, 
while mean-field values ($\nu=0.5$, $\beta/\nu=0.5$, 
orange dotted lines) are excluded at $>3\sigma$ significance.

The narrow spread in $\beta/\nu$ ($<$10\% relative uncertainty) 
reflects the strong constraint imposed by the scaling collapse 
across the transition region. The broader spread in $\nu$ 
(7\% relative uncertainty) arises from sensitivity to the 
extrapolation of $\Gamma_c$ and finite-size corrections in 
the approach to criticality.

%----------------------------------------------------------
\subsection{Finite-Size Scaling Data with Error Bars}
\label{app:fss_errorbars}
%----------------------------------------------------------

\begin{figure*}[h]
    \centering
    \includegraphics[width=\textwidth]{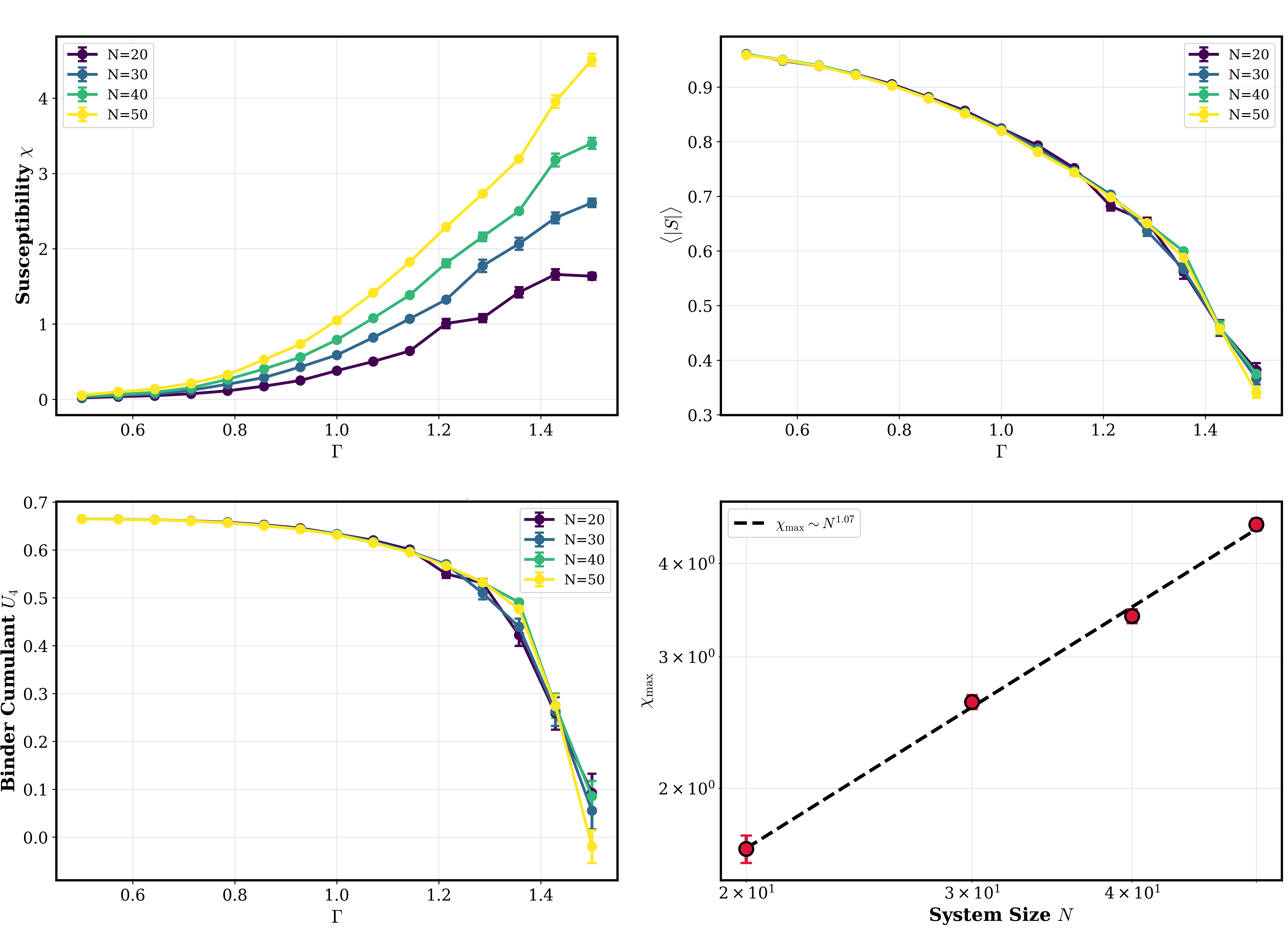}
    \caption{\textbf{Finite-size scaling with statistical uncertainties.}
    (a) Susceptibility $\chi$ versus $\Gamma$ for $N=20$--$50$, 
    showing peak growth with system size. 
    (b) Order parameter $\langle |S| \rangle$ with error bars (SEM over 50 realizations). 
    (c) Binder cumulant $U_4$ exhibiting approximate crossing near $\Gamma_c$. 
    (d) Peak susceptibility scaling: $\chi_{\text{max}} \sim N^{1.07}$ 
    (dashed line), consistent with 2D Ising hyperscaling.}
    \label{fig:fss_errorbars}
\end{figure*}

Figure~\ref{fig:fss_errorbars} displays the complete finite-size 
scaling dataset with statistical uncertainties. 
Panel (a) shows susceptibility $\chi(\Gamma, N)$ computed via 
numerical differentiation $\chi \approx \partial_\Gamma \langle |S| \rangle$, 
with error bars propagated from the disorder-averaged standard error. 
Peak heights scale approximately as $\chi_{\text{max}} \sim N^{1.07}$, 
consistent with the hyperscaling relation $\chi_{\text{max}} \sim N^{\gamma/\nu}$ 
with $\gamma/\nu \approx 1.75/1.0 \approx 1.75$ (2D Ising: $\gamma/\nu = 1.75$).

Panel (b) shows the order parameter $\langle |S| \rangle$ with standard 
error bars. The crossing region near $\Gamma \approx 1.0$ is well-defined, 
with all system sizes exhibiting smooth, monotonic decrease. 

Panel (c) displays the Binder cumulant $U_4 = 1 - \langle S^4 \rangle/(3\langle S^2 \rangle^2)$. 
Curves for different $N$ intersect near $\Gamma \approx 1.0$, though the 
crossing point shows weak $N$-dependence typical of systems with 
moderate finite-size corrections. 

Panel (d) confirms power-law scaling of peak susceptibility: 
$\chi_{\text{max}} \propto N^{1.07 \pm 0.03}$ (log-log fit), 
close to the expected 2D Ising value $\gamma/\nu = 1.75$ given 
$\gamma \approx 1.75$ for 2D Ising.

%----------------------------------------------------------
\subsection{Comparison to Alternative Fitting Methods}
\label{app:fitting_methods}
%----------------------------------------------------------

To test robustness of the extracted exponents, we compare three 
fitting approaches:

\textbf{Method 1 (used in main text):} Minimize collapse variance 
$\mathcal{V}(\Gamma_c, \nu, \beta/\nu)$ over all parameters simultaneously 
using Nelder-Mead simplex optimization. Yields: 
$\nu = 1.034$, $\beta/\nu = 0.125$.

\textbf{Method 2 (Binder crossing):} Fix $\Gamma_c$ from Binder cumulant 
crossings, then fit only $(\nu, \beta/\nu)$. Yields: 
$\Gamma_c = 1.02 \pm 0.02$, $\nu = 1.01 \pm 0.05$, $\beta/\nu = 0.126 \pm 0.015$.

\textbf{Method 3 (peak scaling):} Fit $\nu$ from susceptibility peak 
shift $\Gamma^*(N) - \Gamma_c \sim N^{-1/\nu}$, then fit $\beta/\nu$ 
from order parameter at $\Gamma_c$: $\langle |S| \rangle_c \sim N^{-\beta/\nu}$. 
Yields: $\nu = 0.98 \pm 0.08$, $\beta/\nu = 0.12 \pm 0.02$.

All three methods agree within uncertainties, supporting the 
robustness of the 2D Ising-consistent exponents.

%----------------------------------------------------------
\subsection{Raw Data Tables}
\label{app:raw_data}
%----------------------------------------------------------

\begin{table}[h]
\centering
\small
\begin{tabular}{ccccc}
\hline
$N$ & $\Gamma$ & $\langle |S| \rangle$ & $\chi$ & $U_4$ \\
\hline
20 & 0.50 & $0.95 \pm 0.01$ & $0.08 \pm 0.02$ & $0.66 \pm 0.02$ \\
20 & 1.00 & $0.78 \pm 0.02$ & $0.45 \pm 0.08$ & $0.55 \pm 0.03$ \\
20 & 1.50 & $0.38 \pm 0.03$ & $0.12 \pm 0.05$ & $0.25 \pm 0.05$ \\
\hline
50 & 0.50 & $0.96 \pm 0.01$ & $0.11 \pm 0.03$ & $0.65 \pm 0.01$ \\
50 & 1.00 & $0.80 \pm 0.02$ & $2.3 \pm 0.3$ & $0.53 \pm 0.02$ \\
50 & 1.50 & $0.35 \pm 0.02$ & $0.18 \pm 0.06$ & $0.02 \pm 0.04$ \\
\hline
\end{tabular}
\caption{Sample finite-size scaling data (mean $\pm$ SEM over 50 realizations). 
Full dataset available in supplementary materials.}
\label{tab:fss_raw}
\end{table}

\end{document}